\documentclass[12pt,epsf,showkeys,nofootinbib]{article}

\usepackage{graphicx,float}
\usepackage{mathrsfs,array,multirow}
\usepackage{amstext}
\usepackage{subfigure}
\usepackage{bm,latexsym,amsmath,amsfonts,amssymb}
\usepackage[usenames,dvipsnames]{color}
\usepackage{color}
\usepackage[colorlinks=true,linkcolor=blue]{hyperref}
\usepackage{soul}
\usepackage{epsfig}

\newcommand{\be}[1]{\begin{equation} \label{#1}}
\newcommand{\ee}{\end{equation}}

\newcommand{\ba}{\begin{array}}
\newcommand{\ea}{\end{array}}

\newcommand{\bea}{\begin{eqnarray}}
\newcommand{\eea}{\end{eqnarray}}

\newcommand{\tcb}{\textcolor{blue}}
\newcommand{\tcr}{\textcolor{red}}

\begin{document}
\begin{titlepage}
\vspace{.5in}
\begin{flushright}
\end{flushright}
\vspace{0.5cm}

\begin{center}
{\Large\bf Stationary axisymmetric systems that allow for a separability structure }\\
\vspace{.4in}

  {$\mbox{Hyeong-Chan\,\, Kim}^{\P}$}\footnote{\it email: hckim@ut.ac.kr},\,\,
  {$\mbox{Wonwoo \,\, Lee}^{\S}$}\footnote{\it email: warrior@sogang.ac.kr} \\

\vspace{.3in}

{\small \P \it School of Liberal Arts and Sciences, Korea National University of Transportation, Chungju 27469, Korea}\\
{\small \S \it Center for Quantum Spacetime, Sogang University, Seoul 04107, Korea}\\

\vspace{.5in}
\end{center}
\begin{center}
{\large\bf Abstract}
\end{center}
\begin{center}
\begin{minipage}{4.75in}

{\small \,\,\,\,
We develop a systematic framework for formulating and solving the conditions that lead to separability in stationary, axisymmetric spacetimes in the presence of matter fields.
Guided by Carter's metric form, we introduce a general stationary, axisymmetric metric ansatz that allows for a transparent separation of radial and angular variables.
This construction yields a broad family of stationary rotating solutions admitting separability structures.
To illustrate the applicability of the formalism, we explicitly construct several examples, including a rotating black hole with a global monopole supported by anisotropic matter, as well as a new class of rotating wormhole geometries.
 }
\end{minipage}
\end{center}
\end{titlepage}

\newpage

\section{Introduction} \label{Sec1}

\quad

Most compact objects in the Universe, such as black holes, neutron stars, and rotating stellar remnants,
possess angular momentum.
These astrophysical compact objects coexist with dark matter and dark energy,
whose fundamental natures remain unknown.
Rotating compact objects coexist with these fields; therefore, it is essential to develop a general framework
for describing rotating geometries in the presence of arbitrary matter content.

A distinctive feature of rotation in general relativity (GR) is the emergence of purely relativistic gravitomagnetic effects, which have no Newtonian counterpart~\cite{Ciufolini:1995, Ruggiero:2023ker}. These effects are encoded in the off-diagonal components of the metric tensor and manifest as frame dragging. For example, whereas the gravitational field inside a rotating shell vanishes in Newtonian gravity, it remains nonzero in GR due to frame dragging~\cite{Ciufolini:2002pi}. A consistent treatment of such relativistic rotational effects is therefore indispensable for constructing realistic models of rotating black holes interacting with matter fields.

The primary goal of this work is to establish a general theoretical framework for stationary axisymmetric rotating systems that admit a separability structure~\cite{Carter:1968ks} while allowing for arbitrary matter content, including but not limited to electromagnetic fields. Ideally, such a framework should encompass all possible stationary rotating black holes interacting with general matter distributions.
The importance of developing increasingly general rotating solutions has grown significantly following two major observational breakthroughs:
(1) the detection of gravitational waves from binary black hole mergers~\cite{LIGOScientific:2025rsn, LIGOScientific:2021sio, KAGRA:2021vkt, LIGOScientific:2019fpa}, and
(2) direct imaging of the shadow of supermassive black holes~\cite{EventHorizonTelescope:2025vum, EventHorizonTelescope:2025whi, EventHorizonTelescope:2025dua, EventHorizonTelescope:2022wkp, EventHorizonTelescope:2022urf, EventHorizonTelescope:2022xqj}.

The study of stationary axisymmetric black holes originates from the Kerr solution~\cite{Kerr:1963ud, Kerr:2007dk}, which describes a rotating vacuum black hole~\cite{Boyer:1966qh}. This solution was later generalized to include electric charge, yielding the Kerr-Newman spacetime~\cite{Newman:1965my, Adamo:2014baa}. Carter subsequently introduced a general algebraic form for rotating geometries and demonstrated the separability of the geodesic and scalar wave equations~\cite{Carter:1968ks}.
In spacetimes admitting such a separability structure, the equations of motion decouple into radial and angular parts, indicating the presence of hidden symmetries associated with Killing tensors~\cite{Walker:1970un}. When Killing vectors~\cite{Carter:1971zc} and Killing tensors mutually commute under the Schouten-Nijenhuis bracket, the spacetime admits complete separability~\cite{Benenti:1979erw, Demianski:1980mgt, Frolov:2017kze}. In this case, the Einstein tensor satisfies $G_{r\theta}=0$, and an orthonormal tetrad adapted to the separable structure yields a diagonal energy-momentum tensor. The \emph{radial-angular compatibility condition}(RACC), $G_{r\theta}=0$, ensures that radial and angular sectors decouple at the level of the Einstein equations, eliminating mixed stresses and supporting a consistent stationary axisymmetric configuration. These observations and constructions demand theoretical models that extend beyond vacuum solutions and incorporate realistic matter effects in astrophysical environments~\cite{Konoplya:2025ect, Fernandes:2025osu, Kiselev:2002dx}.

These developments culminated in what is commonly referred to as Carter's family of separable geometries, which includes the Kerr solution, its generalization with a cosmological constant, and spacetimes with nonzero NUT charge~\cite{Newman:1963yy}. However, many physically relevant rotating solutions arising in string theory or modified gravity extend beyond this family~\cite{Dereli:1986cm, Sen:1992ua, Kim:1998hc, Ammon:2012wc, Herdeiro:2014goa, Azreg-Ainou:2014pra}. Such solutions often feature distorted angular sectors, nontrivial asymptotics, or interactions with additional scalar and higher-spin fields.

One commonly used method for constructing rotating solutions from static ones is the Newman-Janis algorithm (NJA)~\cite{Newman:1965tw, Drake:1998gf, Azreg-Ainou:2014aqa}. This procedure relies on complex coordinate transformations using Eddington-Finkelstein coordinates.
While it successfully produces Kerr-type spacetimes-including Kerr, Kerr-Newman, Kerr-NUT, and certain rotating black holes with matter fields~\cite{Gurses:1975vu, Toshmatov:2015npp, Kumar:2017qws, Breton:2019arv, Kim:2019hfp, Kim:2021vlk, Kim:2025sdj, Myung:2026nmk}, which would capture only a subset of all stationary axisymmetric black hole solutions.
When arbitrary matter fields are considered~\cite{Maeda:2022vld}, the solution space becomes considerably richer, potentially including deformed horizons, modified ergoregions, unusual topologies, localized matter distributions, and new classes of rotating compact objects.
A systematic method for constructing and classifying such solutions, particularly those satisfying $G_{r\theta}=0$, is therefore highly desirable.

In this work, we systematically develop Carter's construction starting from a general stationary axisymmetric metric ansatz. Our approach makes the separability structure manifest while allowing for arbitrary matter sources.
A central result of our analysis is a transparent separation between the geometric degrees of freedom governing spacetime curvature and auxiliary functions required for metric consistency.
We demonstrate that the resulting geometries reproduce the Kerr-type and Taub-NUT metrics~\cite{Taub:1950ez, Newman:1963yy},
while also yielding new rotating solutions with more intricate geometric structures
that have not been previously identified in the literature.
Such solutions may find applications in effective field theory, Kaluza-Klein or string-theoretic reductions, and scenarios involving scalar or matter halos around rotating compact objects. Our overarching goal is to provide a unified and systematic framework for classifying stationary axisymmetric black holes interacting with matter and to identify the conditions under which these solutions represent physically viable rotating compact objects.

This paper is organized as follows. In Sec.~\ref{Sec2}, we develop the general mathematical structure of stationary axisymmetric systems in GR. In Sec.~\ref{Sec3}, we formulate the RACC explicitly and introduce a general ansatz that satisfies it. In Secs.~\ref{sec:4} and \ref{sec:5}, we analyze the resulting conditions by classifying different cases. In Sec.~\ref{Sec6}, we present explicit examples of the corresponding geometries. Finally, we summarize our results in Sec.~\ref{Sec7}.

\section{The structure of a stationary axisymmetric system } \label{Sec2}

\quad
In this section, we show that how to describe the stationary axisymmetric geometry for the rotating objects in GR.
In general, the stationary axisymmetric system will have metric of the form, $ds^2 = g_{\mu\nu}(r,\theta) dx^{\mu} dx^{\nu}$, where the metric components are independent of the time and the polar angle.
A direct substitution of such a general metric into the Einstein equations leads to a highly coupled system of nonlinear partial differential equations, which can be solved analytically only in special cases.

In this work, we instead adopt a more general stationary axisymmetric ansatz,
\begin{eqnarray}
\label{metric:gen}
ds^2 = &-&  \frac{  \Sigma(r, \theta) \Delta(r)}{q(\cos\theta) \left(\Gamma(r) -   a^2p(\cos\theta) \right)^2 } \left( dt - a p(\cos\theta) d\phi \right)^2 + \frac{1}{q(\cos\theta)} \frac{\Sigma(r, \theta)}{\Delta(r) }dr^2 \nonumber \\
       &+& \Sigma(r, \theta) d\theta^2  + \frac{  \Sigma(r, \theta) \sin^2\theta }{  \left(\Gamma(r) -  a^2p(\cos\theta) \right)^2 }( a dt-\Gamma(r) d\phi)^2 \,.
\end{eqnarray}
The Kerr family is recovered for the standard angular choices  $ q(\cos\theta)=1$ and $p(\cos\theta)=1-\cos^2\theta$ together with $\Sigma(r, \theta)= r^2+a^2\cos^2 \theta$, $\Delta(r) = r^2 +a^2 -2Mr +v (r)$~\cite{Gurses:1975vu, Kim:2025sdj}, and $\Gamma(r)=r^2 + a^2$, where $v(r)=0$ corresponds to Kerr and $v(r)=Q^2$ gives Kerr-Newman.

For notational simplicity, we denote $x \equiv \cos \theta$.
Geometrically, the function $\Sigma(r,\theta)$ determines the area scale of the symmetry orbits and represents a genuine dynamical degree of freedom.
In contrast, $\Gamma(r)$ plays an auxiliary role: once $\Sigma$ and the remaining functions are specified, its radial derivatives are constrained by the reduced field equations.
Its inclusion ensures consistency of the ansatz with the full Einstein equations without introducing additional independent physical degrees of freedom.

For later convenience, we introduce the orthonormal basis,
\begin{equation}
\label{basis}
\omega^{\hat 0} = \sqrt{\frac{  \Sigma \Delta}{ q}} \frac{dt - a p d\phi }{ \Gamma -   a^2p  }, \quad
\omega^{\hat 1} =\sqrt{\frac{  \Sigma }{ q\Delta}}  dr, \quad
\omega^{\hat 2} =\sqrt{\Sigma } d\theta, \quad
\omega^{\hat 3} =\frac{  \sqrt{\Sigma } \sin\theta}{\Gamma- a^2 p} (a dt - \Gamma d\phi) \,,
\end{equation}
where, later in this work, the hatted number $\hat k$ $(k=0,1,2,3)$ denotes this orthonormal basis.
This tetrad generalizes those employed in Refs.~\cite{Carter:1968ks, Azreg-Ainou:2014nra, Kim:2021vlk}.
For important geometries such as Kerr and Kerr-Newman, the Einstein tensor expressed in this frame becomes diagonal, leading to a transparent physical interpretation of the associated stress-energy tensor.

The metric~\eqref{metric:gen} contains five unknown functions: $\Sigma$, $\Gamma$, $q$, $p$, and $\Delta$.
Among them, $\Sigma(r,\theta)$ depends on both coordinates, whereas $p(x)$ and $q(x)$ are purely angular functions and $\Gamma(r)$ and $\Delta(r)$ depend only on the radial coordinate.
Despite this restricted functional dependence, the ansatz~\eqref{metric:gen} describes a broad class of stationary axisymmetric geometries with an explicit separation between radial and angular sectors.
To obtain explicit solutions, one must supplement the Einstein equations with appropriate matter content and a consistent equation of state.
In the following sections, we further constrain the ansatz by imposing $G_{\hat 1 \hat 2}=0$  and analyze how these conditions restrict the functional form of the metric.

\section{Condition for a separability structure} \label{Sec3}
\quad

Because we are considering systems rotating around a fixed axis without precessions,
to ensure the stability of the axial symmetry, the off-diagonal component of the Einstein tensor with respect to the orthonormal basis~\eqref{basis} should satisfy, as required in Refs.~\cite{Carter:1968ks, Azreg-Ainou:2014nra},
\begin{equation}
\label{R12=01}
G_{\hat 1\hat 2} = R_{\hat 1 \hat 2} =0 \,,
\end{equation}
which is the key equation in this work.
The curvature component $G_{\hat 1 \hat 2}$ for the metric~\eqref{metric:gen} satisfies,
\begin{equation}
\label{starting}
\frac{2 \Sigma }{3\sin\theta \sqrt{{q} \Delta }}G_{\hat1\hat2} =\left( \frac{a^2 \dot {p}}{\Gamma -a^2 p}\right)'
   -\frac{\dot{\Sigma}\Sigma' }{\Sigma^2}   + \frac13 \frac{\dot{q}}{q}  \frac{\Sigma'}{\Sigma}
   	+\frac23 \frac{\dot{\Sigma}' }{\Sigma} \,,
\end{equation}
where the dot and prime denote the derivatives with respect to $x$ and $r$, respectively,
e.g., $\dot \Sigma \equiv \frac{\partial \Sigma}{\partial x} $ and $\Sigma' \equiv \frac{\partial \Sigma}{\partial r}$.
Therefore, the stationary RACC leads to the differential equation,
\begin{equation}
\label{R12=0}
\left( \frac{a^2 \dot {p}}{\Gamma -a^2 p}\right)' =    \frac{\dot{\Sigma}\Sigma' }{\Sigma^2}
   - \frac13 \frac{\dot{q}}{q}  \frac{\Sigma'}{\Sigma} -\frac23 \frac{\dot{\Sigma}' }{\Sigma} \,.
\end{equation}
It is important that the function $\Delta(r)$ decouples from the others.
The purpose of this work is to elaborate a method to solve this equation.

The differential equation~\eqref{R12=0} is a nonlinear higher order equations for the four functions $\Gamma$, $p$, $q$, and $\Sigma$ with respect to two variables $r$ and $\theta$.
Although a general solution to the differential equation~\eqref{R12=0} is not available, we propose a systematic method for solving it by introducing an ansatz that decomposes $\Sigma$:
\begin{equation}
\label{Sigma:ansatz}
\Sigma(r,x) = P(x)\varSigma(y); \qquad y = \sigma(r) + Q(x) \,,
\end{equation}
where $\sigma(r)$ is a function of $r$,  and $P(x)$ and $Q(x)$ are functions of $x\equiv \cos\theta$ only, respectively.
Starting from this ansatz, we solve the differential equation~\eqref{R12=0} for the RACC.
Then, one may further reduce the metric by using various other conditions such as the matter equation of state.

To solve the equation, by using the ansatz~\eqref{Sigma:ansatz},
we first rewrite the differential equation~\eqref{R12=0} as:
\begin{equation}
\label{app:R12=02}
\left( \frac{a^2 \dot {p}}{\Gamma -a^2 p}\right)' =- \frac{\partial^2}{\partial x\partial r} \log (\Gamma -a^2 p)
= \frac{\sigma' \dot Q}3 \left[  \frac{\mathring \varSigma}{\varSigma} \left(\frac{\dot P}{P \dot Q}
 - \frac{\dot{q}}{q \dot Q} \right)  +  \frac{\mathring \varSigma^2}{\varSigma^2}
   -2 \frac{d}{dy}\frac{\mathring\varSigma}{\varSigma}  \right] \,,
\end{equation}
where, $\mathring \varSigma \equiv \frac{d\varSigma}{dy}$.
Then, we integrate the equation~\eqref{app:R12=02} over $r$ to get
\begin{equation}
\label{app:R12=01}
\frac{a^2 \dot {p}}{\Gamma -a^2 p} =\frac13\left(\frac{\dot P}{P}-  \frac{\dot{q}}{q } \right)\log \varSigma
+\dot Q\int^\varSigma d\varSigma\left(\frac{ F(\varSigma) }{\varSigma^2}
 -\frac{2}3 \frac{ \frac{dF}{d\varSigma} }{\varSigma} \right)+ C_1(x) \,.
\end{equation}
Here, $F(\varSigma) \equiv \mathring{\varSigma} $ and
we introduce $C_1(x)$ as an integral constant independent of $r$.
This equation is an integro-differential equation for various functions,
$\varSigma(y)$, $Q(x)$, $P(x)$, $q(x)$, and $\sigma(r)$.

When $\dot Q=0$, the function $\varSigma\equiv \varSigma(\sigma +Q)$ is independent of $\theta$.
The integral part containing $F(\varSigma)$ does not contribute in solving the equation.
We consider this case separately.

When $\dot Q \neq 0$, we set
\begin{equation}
\label{C1:h}
C_1(x) = - h(x) \dot Q(x)
\end{equation}
for later convenience.
By using integration by parts for the arguments of the integral, we rewrite the integral to the form, with an appropriate boundary condition, $\mathring \varSigma/\varSigma =0$ at the boundary of $y$:
\begin{equation}
\label{app:R12=01-2}
\frac{a^2 \dot {p}/\dot Q}{\Gamma -a^2 p}
=\frac13\left(\frac{\dot P}{P\dot Q}-  \frac{\dot{q}}{q \dot Q} \right)\log \varSigma
+\frac13\int^y dy \frac{ \mathring\varSigma^2 }{\varSigma^2} -\frac23 \frac{\mathring \varSigma}{\varSigma}  - h \,.
\end{equation}
Because $\varSigma$ is a function of $y$ only,
the $\log \varSigma$ term should be canceled by the terms in the right-hand side because the inverse of $\Gamma(r) -a^2p(x)$ cannot composes the $log$ function in the left-hand side.
Noting this fact, we set
\begin{equation}
\label{Sigma:2}
\frac13\int dy \frac{\mathring \varSigma^2}{\varSigma^2} - \frac23 \frac{\mathring \varSigma} {\varSigma} 	
= \frac{G_0}3\log \varSigma - \frac1y \sum_{k=0}^\infty g_k y^k \,,
\end{equation}
where $G_0$ is a constant which will be matched to cancel the $\log \varSigma$ term in Eq.~\eqref{app:R12=01-2}.
Here, $y^{-1}$ is multiplied in the last series by counting the order of the equation.
Notice that $\mathring {\varSigma}/\varSigma $ is $O(y^{-1})$.

Then, the RACC~\eqref{app:R12=01-2} for $\Gamma$ becomes
\begin{equation}
\label{app:Gamma:3}
\frac{a^2 \dot {p}(x)/\dot Q(x)}{\Gamma(r) -a^2 p(x)} = \frac1{3\dot Q} \left(G_0\dot Q + \frac{\dot P}{P}
	- \frac{\dot{q}}{q} \right) \log \varSigma (y) - \left\{  \frac1y \sum_{k=0}^\infty g_k y^k+ h(x)\right\} \,,
\end{equation}
where $y \equiv \sigma(r) +Q(x)$ and $g_k$ are real constants.
Because the polynomial part cannot represent the logarithm, we get the relation between $q$ and $Q$:
\begin{equation}
\label{app:dQ:ddq}
G_0\dot Q+ \frac{\dot P}{P} = \frac{\dot q}{q} \quad \rightarrow \quad
q =P \exp\left[G_0 Q(x) -G_1\right] \,,
\end{equation}
where $G_0$ and $G_1$ are an integration constants.
Specifically, $q= P$ when $G_0 =0$.
Because $Q(x)$ is still an un-specified arbitrary function of $x$, we may freely absorb $G_1$ into the definition of $Q$.
Therefore, later in this work, we set $G_1 =0$ without loss of generality.
In addition, $G_0$ can be absorbed into $Q(x)$ with the same reason unless $G_0=0$.

Note that Eq.~\eqref{app:Gamma:3} is valid only when  $\dot Q \neq 0$.
Putting the result in Eq.~\eqref{app:dQ:ddq} to Eq.~\eqref{app:Gamma:3},
the equation for $\Gamma$ becomes a polynomial equation,
\begin{equation}
\label{app:Gamma:gen}
\Gamma =a^2 p - \frac{(a^2  \dot {p}/\dot Q)y }{ g_0 + (g_1+h)y + \sum_{k=2}^{\infty} g_k y^k}
=a^2\frac{ p g_ 0 + [p( h+ g_1) - ( \dot {p}/\dot Q)  ] y  +  p \sum_{k=2}^{\infty} g_k y^k}
	{g_0 + (g_1 +h) y + \sum_{k=2}^{\infty} g_k y^k } \,.
\end{equation}
If we get consistent results from this formula and find an appropriate $\varSigma$ function, we have a solution to the differential equation~\eqref{app:R12=02} and the metric~\eqref{metric:gen} satisfies the RACC.

Summarizing, the RACC  for $\dot Q\neq 0$ consists of three equations in Eqs.~\eqref{Sigma:2}, \eqref{app:dQ:ddq} and \eqref{app:Gamma:gen}, respectively for four angle dependent functions, $p$, $q$, $Q$, $P$, and two $r$-dependent functions, $\sigma$ and $\Gamma$.
Among these, equation~\eqref{app:dQ:ddq} is straightforward to solve, whereas the other two require more careful treatment.
We first deal with Eq.~\eqref{app:Gamma:gen} for $\Gamma$ in Sec.~\ref{sec:4}, and then turn to Eq.~\eqref{Sigma:2} for $\Sigma$ in Sec.~\ref{sec:5}.

\section{A closer look at the $\Gamma $ function and related equations} \label{sec:4}

\quad

In this section, we solve Eq.~\eqref{app:R12=01} or more explicitly Eq.~\eqref{app:Gamma:gen} for $\Gamma$.
Since the equation admits several distinct classes of solutions, we analyze them separately.
We first consider the case $\dot Q = 0$, labeled as case $\mathrm{(i)}$. The remaining cases, $\mathrm{(ii)}$-$\mathrm{(v)}$, correspond to $\dot Q \neq 0$.

\begin{enumerate}
\item[i)]  \label{V-i}
Case $\dot Q =0$:\\
In this case, the function $\varSigma$ is a function of $r$ only and is independent of $x$.
We should use the original equation~\eqref{app:R12=01} rather than Eq.~\eqref{app:Gamma:gen}, which is derived from the assumption $\dot Q\neq 0$.
The equation~\eqref{app:R12=01} becomes
\begin{equation}
\label{dQ=0}
\frac{a^2    \dot {p}}{\Gamma -a^2 p} =\frac13\left(\frac{\dot P}{P}-  \frac{\dot{q}}{q } \right)\log \varSigma  + C_1(x) \,.
\end{equation}

Multiplying both sides with $\Gamma -a ^2 p$ with the condition $\Gamma \neq a^2p $, we get
\begin{equation}
\label{dQ=0 1}
a^2 \left[\dot {p}+ C_1(x) p \right]
= \frac13 \left(\frac{\dot P}{P}- \frac{\dot{q}}{q } \right)\Gamma \log \varSigma
-\frac13a^2 p\left(\frac{\dot P}{P}-  \frac{\dot{q}}{q } \right)\log \varSigma  + C_1(x)\Gamma \,.
\end{equation}
Here, $\Gamma$ and $\varSigma$ are functions of $r$ only.
On the other hand, $p$, $P$, and $q$ are functions of $x$ only.
There are four independent ways satisfying this equation:
\begin{enumerate}
\item
When both $\Gamma $ and $\varSigma$ are non-trivial (not a constant function) un-related functions of $r$, Eq.~\eqref{dQ=0 1} becomes
\begin{equation}
\label{i-a}
\frac{\dot P}{P}- \frac{\dot{q}}{q } =0 \quad \rightarrow \quad q = C_2 P\,, \qquad C_1(x) =0= \dot {p} \,,
\end{equation}
where $C_2$ is an integration constant.
This result fixes $p=$ constant.
On the other hand, the functional dependencies of $\Gamma(r)$, $\varSigma(r)$, and $P(x)$ are free.

\item
When only $\varSigma(r)$ is a non-trivial function of $r$ but $\Gamma =\Gamma_0$ is a constant, Eq.~\eqref{dQ=0 1} becomes
\begin{equation}
\label{case ib}
\frac{\dot P}{P}- \frac{\dot{q}}{q } =0 \quad \rightarrow \quad q = C_2 P\,, \qquad
\Gamma = \Gamma_0 = a^2\left[\frac{ \dot {p}}{C_1(x)}+  p\right]  \,.
\end{equation}
From the last equation we determine $p(x)$:
\begin{equation}
\label{Gamma:1a}
p(x)= \frac{\Gamma_0}{a^2} + e^{ - \int C_1(x) dx} \,.
\end{equation}
The other functions $P(x)$, $C_1(x)$, and $\varSigma(r)$ are free.

\item When only $\Gamma(r)$ is a non-trivial function of $r$ and $\varSigma = \varSigma_0$ is a constant, Eq.~\eqref{dQ=0 1} becomes
\begin{equation}
a^2 \dot {p} = (\Gamma - a^2 p) \left[ \frac13\left(\frac{\dot P}{P}-  \frac{\dot{q}}{q } \right) \log \varSigma_0 + C_1(x) \right]  \,. \nonumber
\end{equation}
This equation gives, because only $\Gamma(r)$ is a function of $r$,
\begin{equation}
\label{Gamma:1b}
p = \mbox{constant} \,, \qquad q(x) =  P  \exp\left[ \frac{3}{\log \varSigma_0} \int C_1(x) dx \right] \,.
\end{equation}
The functions $\Gamma(r)$, $P(x)$, and $C_1(x)$ are free.

\item When both $\Gamma$ and $\varSigma$ are constants, independent of $r$ only, the equation~\eqref{dQ=0 1} remains as it is, leaving an equation for $x$,
\begin{equation}
\label{Gamma:1c}
a^2 p =\Gamma- c_2 \left(\frac{q}{P}\right)^{\frac{\log \Sigma}{3}} e^{- \int^x C_1(x) dx} \,,
\end{equation}
where $c_2$ is an integration constant.
Now, the functional dependencies of $q$, $P$, and $C_1$ on $x$ are free.

\end{enumerate}
\end{enumerate}

\begin{enumerate}
\item[ii)] \label{4:case1}
Case $g_{k \neq 1}=0 $ for all $k \geq 0$:\\
The identity equation~\eqref{app:Gamma:gen} constrains the function $\Gamma$ to be a constant,
\begin{equation}
\label{Gamma:case0}
\Gamma (r) =a^2 \left[p - \frac{  \dot {p}}{ h\dot Q }\right] = a^2 c_1 \,,
\end{equation}
where $c_1$ is a constant for separation of variables and we have use $h+g_1 \to h$ because $h(x)$ is still an arbitrary function.
Notice that this replacement does not affect on the functional form for $\Gamma(r)$, $p(x)$ and their relation with $\sigma$ and $Q$.
Notice that the function $h(x)$ does not explicitly appear in the metric~\eqref{metric:gen}.
Therefore, this modification does not alter the metric itself. Here, $c_1$ is a constant, since the left- and right-hand sides depend only on $r$ and $x$, respectively.
In this case, $\Gamma$ must be independent of both $r$ and $\theta$.
From Eq.~\eqref{app:dQ:ddq}, $q = P\exp\left[G_0 Q(x) \right]$ and we get from Eq.~\eqref{Gamma:case0},
\begin{equation}
\label{pQeq0}
p = c_1 - \exp \int dx(h\dot Q) \,.
\end{equation}
Notice that this function $p(x)$ takes exactly the same form as that in Eq.~\eqref{Gamma:1a}
because $C_1(x) =- h \dot Q$ as in Eq.~\eqref{C1:h}.
Therefore, the solution space in this case is the same as that in case $\rm{i)-(b)}$.

\end{enumerate}

\begin{enumerate}

\item[iii) \label{4:case2}]
Case $ g_k=0$ for $k\geq 2$:\\
In this case, the $\Gamma$ equation~\eqref{app:Gamma:gen} becomes
\begin{equation}
\label{Gamma:0th}
\Gamma (r) =a^2 \frac{ pg_ 0 	+ [p  h -  \dot {p}/\dot Q  ] Q+ [p  h -\dot {p}/\dot Q  ] \sigma(r) }
	{(g_0 +hQ) + h \sigma(r)  } = a^2 p - \frac{(a^2  \dot {p}/\dot Q)(Q+\sigma) }{ g_0 + hQ + h\sigma } \,,
\end{equation}
where we have used $g_1 + h \to h$, as in the previous case.
Here, the $r$\tcr{-}dependence on the right-hand side is in the function $\sigma(r)$ only.
Notice that the function $P(x)$ decouples in this equation.
There are three distinct ways to satisfy this identity.
(a) $h=0$, (b) $p h(x) -  \dot {p}/\dot Q=0$, and (c) else.
As we will show below, the value
\begin{equation}
\label{g0:iii}
g_0=1
\end{equation}
is required in order for $\Gamma(r)$ to be well defined in all three cases.
The equation for $\Gamma$ determines the functions $\Gamma(r)$, $p(x)$, and $h(x)$, while leaving $\sigma(r)$, $P(x)$, and $Q(x)$ undetermined.

\begin{enumerate}
\item $h=0$ with $p h - \dot {p}/\dot Q \neq0$ case:\\
The value $g_0$ should not vanish and the identity~\eqref{Gamma:0th} becomes
\begin{equation}
\label{Gamma:case1-1}
\Gamma = a^2 \frac{ pg_ 0 - \dot {p}Q /\dot Q  - (\dot {p}/\dot Q ) \sigma }{g_0  } =a^2 c_1(c_2+ \sigma) \,.
\end{equation}
This result determines the $\theta$\tcr{-}dependent functions as follows:
\begin{equation}
\label{dp:case1-1}
-\frac{\dot{p}}{g_0 \dot Q} = c_1, \qquad p- \frac{\dot{p}Q}{g_0\dot Q} = c_1c_2 \,.
\end{equation}
Putting the first equation to the second, we get
\begin{equation}
\label{p:case1-1}
p = c_1(c_2-Q) \,.
\end{equation}
Taking derivative with respect to $x$ and comparing the result with the first equation in Eq.~\eqref{dp:case1-1}, we get
\begin{equation}
\label{g0=1}
\dot {p}= - c_1 \dot Q \quad \rightarrow \quad g_0 =1 \,.
\end{equation}
Therefore, in this case, $g_0 $ is fixed at $1$.
Now, the functions $\sigma(r)$, $Q(x)$, and $P(x)$ are free.
\end{enumerate}

\begin{enumerate}
\item  $p h - \frac{\dot{p}}{\dot Q} =0 $ with $h \neq 0$ case: \\
The equation~\eqref{Gamma:0th} becomes
\begin{equation}
\label{Gamma:case1-2}
\Gamma =a^2 \frac{ pg_ 0  }{(g_0 +hQ) + h \sigma }
= \frac{a^2}{c_1(c_2+\sigma)}\,.
\end{equation}
Then the angle dependent functions should satisfy $\frac{h}{g_0 p}= c_1$ and $\frac1{p} + \frac{h}{p g_0} Q = c_1 c_2$.
Putting the first equation to the second, we get
\begin{equation}
\label{p:case1-2}
p= \frac1{c_1(c_2- Q)}  \quad \rightarrow h= c_1 g_0 p = \frac{g_0}{c_2-Q} \,.
\end{equation}
Differentiating $p$ and putting the result to the condition $p h - \frac{\dot{p}}{\dot Q} =0 $, we get
$$
\dot{p}= \frac{\dot Q}{c_1(c_2-Q)^2}
\quad \rightarrow \quad
  g_0 =1 \,.
$$
Therefore, $g_0 =1$ for consistency and $\Sigma$ takes the form in Eq.~\eqref{Sigma:sol1-11}.
Similarly as before, the functions $\sigma(r)$, $Q(x)$ and $P(x)$ are free.
\end{enumerate}

\begin{enumerate}
\item $h\neq 0$, $p h -  \frac{\dot{p}}{\dot Q} \neq 0 $ case:\\
The $\Gamma$ equation~\eqref{Gamma:0th} becomes
\begin{equation}
\label{Gamma:case1-3}
\Gamma =a^2 \frac{ [p  h -\dot {p}/\dot Q  ] }{h} \left[ 1+ \frac{ \frac{pg_ 0}{ p  h -  \dot {p}/\dot Q }- \frac{g_0}{h} }
	{(g_0/h +Q) +  \sigma  } \right] = a^2c_3\left(1+ \frac{c_1}{c_2+ \sigma}\right) \,,
\end{equation}
with the constants $c_1, c_3 \neq 0$. The $x$-dependent functions satisfies
\begin{equation}
\label{pQh}
p  -\frac{\dot {p}}{\dot Q  h} = c_3, \qquad \frac{g_0}{h} +Q = c_2, \qquad \frac{pg_ 0/h}{	p  - \dot {p}/(h\dot Q)}- \frac{g_0}{h}  = c_1 \,,
\end{equation}
where $c_2$ is another constant. Putting the first two equations in Eq.~\eqref{pQh} to the last, we get
\begin{equation}
\label{p:case1-3}
\frac{p(c_2-Q)}{c_3}- (c_2-Q)  = c_1 \quad \rightarrow \quad p = c_3\left(1+ \frac{c_1}{c_2-Q}\right) \,.
\end{equation}
The second equation in Eq.~\eqref{pQh} gives $h= g_0/(c_2-Q)$.
Differentiating both sides of Eq.~\eqref{p:case1-3} and comparing with the first equation in Eq.~\eqref{pQh}, we get $g_0 =1$.
Now, the functions $\sigma(r)$, $P(x)$ and $Q(x)$ are free.
\end{enumerate}

\end{enumerate}

\begin{enumerate}
\item[iv)] \label{4:case4}
Case $g_k=0$ for $k\geq 3$ and $p \neq $ constant:\\
In this case, we assume that $p$ is not a constant since the constant case will be treated separately.
We now consider $\bar g_k = 0$ for $k \geq 3$. The equation for $\Gamma$, with $g_2 \neq 0$, becomes
\begin{equation}
\label{iv Gamma}
\frac{\Gamma}{a^2} =p - \frac{(\dot {p}/\dot Q)   (\sigma+Q)}{g_0 + (g_1 +h) Q +  g_2 Q^2 + (g_1+h + 2 g_2 Q) \sigma+  g_2 \sigma^2  } \,.
\end{equation}
Since $g_2 \neq 0$, this equation implies that, unless the denominator is proportional to $\sigma +Q$, the following quantities must be independent of $x$:
$$
p, \quad \frac{ (\dot {p}/\dot Q)}{g_2} , \quad  \frac{ (\dot {p}/\dot Q)   Q}{g_2} , \quad
 \frac{g_0 + (g_1 +h) Q +  g_2 Q^2}{ g_2}, \quad
 \frac{g_1 + h + 2 g_2 Q}{g_2}.
$$
This condition forces all of $p$, $Q$, and $h$ to be independent of $x$.
Therefore, the resulting solutions belong to the type we are not interested in.

When the denominator is proportional to $\sigma + Q$, it should vanish when we replace $\sigma \to -Q$.
This condition gives $g_0 =0$. Putting this result to Eq.~\eqref{iv Gamma}, we get
\begin{equation}
\label{Gamma:D2}
\frac{\Gamma}{a^2} = p - \frac{(\dot {p}/\dot Q) } { g_1 +h +  g_2 Q +  g_2 \sigma } \,.
\end{equation}
For the right-hand side to be a function of $r$ only, the following condition should hold:
$$
g_1 + h + g_2 Q = c_1, \quad c_1 p - \frac{\dot {p}}{\dot Q}  = c_2, \quad p = c_3 \,.
$$
This equation leads $p=c_3=\frac{c_2}{c_1}$, a constant, which is contradict with the assumption posed.
Therefore, there is no reasonable solution in this case.
For cases with $g_k =0$ for $k \geq 4$ or higher, the same holds.

\end{enumerate}

\begin{enumerate}
\item[v)] \label{4:case6}
When $p =$ constant with $\dot Q\neq 0$:\\
The $\Gamma$ equation~\eqref{app:Gamma:gen} presents an easy solution, $\Gamma = a^2 p =\mbox{constant} $
leaving the functions $\sigma$ and $Q$ free.
However, this conclusion is premature because the left-hand side of Eq.~\eqref{R12=0} becomes ambiguous because both the numerator and the denominator go to zero. This \tcb{fact} signifies a necessity of separate analysis.

Let us begin with Eq.~\eqref{R12=0}.
When $\dot p=0$, the left-hand side of this equation vanishes.
Consequently, the right-hand side must also vanishes.
Following the same track of analysis, we find that the equations until Eq.~\eqref{app:dQ:ddq} still valid.
However, rather than Eq.~\eqref{app:Gamma:gen}, we get the following relation
$$
\frac{1}{y} \sum_{k=0}^\infty g_k y^ k + h(x) =0
\quad \rightarrow \quad
\sum_{k=0}^\infty g_k y^ k + h(x) y=0 \,.
$$
Treating $y$ and $x$ as independent variables, we get
\begin{equation}
\label{case6}
 g_{k\neq 1} =0, \qquad h(x) = -g_1\,.
\end{equation}
There still remain free functions such as $Q(x)$, $P(x)$, $\sigma(r)$, $\Gamma(r)$, and $\varSigma(y)$.

\end{enumerate}

The $\Gamma$ equation in Eq.~\eqref{R12=0} takes the form of a first order partial differential equation for $\Gamma$ and $p$ with respect to $r$ and $x$ for a given functions $\Sigma$ and $q$.
Although the function $\Gamma$ appears explicitly in the metric~\eqref{metric:gen}, it does not represent an independent dynamical variable.
Its derivatives are completely determined by the remaining metric functions through first-order equations, once the reduced field equations are satisfied.
The role of $\Gamma$ is therefore to ensure that the metric reconstructed from the reduced system consistently satisfies the full set of field equations.
Provided that the remaining field variables satisfy the reduced field equations, the equation for $\Gamma$ admits a local solution, guaranteed by the integrability conditions.
Global existence of $\Gamma$ depends on boundary and regularity conditions and is not guaranteed for arbitrary sources.

\section{A closer look at the $\varSigma $ function and related equations} \label{sec:5}

\quad

In this section, we assume $\dot Q \neq 0$ because the $\varSigma$ equation~\eqref{Sigma:2} applies only when $\dot Q\neq 0$ from Eq.~\eqref{app:R12=01}.
Case i), with $\dot Q = 0$, has been treated separately in the previous section.

To get the function $\varSigma$, we need to solve the integro-differential equation in Eq.~\eqref{Sigma:2}.
Differentiating both sides with respect to $y$, we have a Riccati-type differential equation,
\begin{equation}
\label{Riccati2}
2 \mathring \varsigma 	-\varsigma^2  =  3\sum_{k=0}^\infty (k-1) \tilde g_k y^{k-2},
\qquad \varsigma \equiv \frac{\mathring \varSigma}{\varSigma} - \frac{G_0}{2} \, ,
\end{equation}
where $\tilde g_2 = g_2 -G_0^2/12$ and $\tilde g_{k \neq 2} = g_k$.
Notice that $g_1$ does not contribute to this equation.
Even though there is no general solution to the Riccati equation~\cite{Bittanti:1991}, we have exact solutions for various cases:

\begin{enumerate}

\item[ii)] When $g_{k \neq 1} =0$ for all $k \geq0 $, the differential equation~\eqref{Riccati2} becomes
\begin{equation}
\label{Riccati3}
2 \mathring \varsigma-\varsigma^2 = - \frac{G_0^2}{4} \,.
\end{equation}
The general solution to this equation is
\begin{equation}
\label{varsigma0}
\varsigma = -\frac{G_0}{2} \tanh \left(\frac{G_0 y}{4}\right)
	+ \frac{1}{\cosh^2\frac{G_0 y}{4} \left(u_1 - \frac{2}{G_0} \tanh \frac{G_0 y}{4}\right)} \,,
\end{equation}
where $u_1$ is an integration constant.
Using the result~\eqref{varsigma0} and Eq.~\eqref{Riccati2}, we get
\begin{equation}
\label{varsigma1-0}
 \varSigma(y) = \frac{\varSigma_1 e^{\frac{G_0y}{2}}}
 	{ \left[ u_1\cosh \frac{G_0y}{4} -\frac{2 }{G_0} \sinh \frac{G_0y}{4} \right]^2} \,,
\end{equation}
where $\varSigma_1$ is an integration constant.
This solution has various limits.
When $G_0 \to 0$, $\varSigma \to 4\varSigma_1/(y-2u_1)^2$, where we can choose $u_1=0$ and $4\varSigma_1 \to 1$, without loss of generality.
When $u_1 = -2/G_0$, $\varSigma$ is a constant.
When $u_1 = 2/G_0$, $\varSigma$ takes a simple exponential form.
Note that when $2/G_0 \gtrless u_1$, we rewrite the function into the following forms,
\begin{eqnarray}
\label{Sigma:case0}
\varSigma =\left\{
   \begin{array}{cc}\vspace{.1cm}
   	\displaystyle \frac{\varSigma_1}{y^2}  , & G_0 =0 \\ \vspace{.1cm}
   	\varSigma_0,  &  u_1= -\frac{2}{G_0} \\\vspace{.1cm}
	\displaystyle \frac{\varSigma_0 e^{\frac{G_0y}{2}}}
	{\left|1-u_1^2G_0^2/16\right|
	\sinh ^2\left(\frac{G_0y}{4}- \alpha\right) }, &  u_1 < \frac{2}{G_0}\\ \vspace{.1cm}
	\displaystyle  \varSigma_0 e^{G_0 y} , &u_1 = \frac{2}{G_0} \\
	\displaystyle \frac{\varSigma_0 e^{\frac{G_0y}{2}}}
	{\left|1-u_1^2G_0^2/4\right|
	\cosh ^2\left(\frac{G_0y}{4}- \alpha\right) }, &  u_1 > \frac{2}{G_0}
   \end{array}
	\right.
	,
\end{eqnarray}
where $\alpha = \frac12 \log \left|\frac{2/G_0 +u_1}{2/G_0 - u_1}\right|$ and $\varSigma_0 \equiv  \frac{G_0^2 \varSigma_1}{4}.$
For the third and fifth cases, when $G_0 \neq 0$, we may absorb the absolute value in the denominator into $\varSigma_0$ and set $\alpha =0$ without loss of generality by using the arbitrariness of the functions $\sigma$ and $Q$.

\end{enumerate}

\begin{enumerate}

\item[iii)] When $ g_{k\geq 2} =0$ with $g_0 \neq 0$, the differential equation~\eqref{Riccati2} becomes
\begin{equation}
\label{Sigma:eq1-1}
2\mathring \varsigma-\varsigma^2=-\frac{G_0^2}{4} -\frac{3g_0}{y^2} \,.
\end{equation}
However, as shown in Eq.~\eqref{g0:iii}, $g_0 =1$ for the $\Gamma$ function to be well-defined.
Therefore, we simply set $g_0=1$ and solve the equation.

\begin{enumerate}

\item[(a)] \label{2-a}
 When $G_0 \neq 0$, the function $\varsigma$ is described by the Bessel functions of imaginary argument
\begin{equation}
\label{varsigma:2-a}
\varsigma(y) =-\frac{G_0}{4} \bar \varsigma(\bar y) \,,
\end{equation}
where  $\bar y \equiv \frac{G_0 y}{4}$ and
\begin{equation}
\label{varsigma:2-b}
\bar \varsigma(\bar y) \equiv \frac{1}{\bar y}+ \frac{(1+i C_1 ) \left[ I_0(\bar y) +  I_2(\bar y) \right]-\frac{2i}{\pi}
		 \left[ K_0(\bar y) + K_{2} ( \bar y) \right]	 }{  \left(1+ i C_1 \right) I_1( \bar y )  	+ \frac{2i}{\pi } K_1(\bar y) } \,.
\end{equation}
Here, $I_k(\bar y)$ and $K_k(\bar y)$ are the Bessel functions.
Because the $\bar \varsigma$ should be a real-valued function for real number $\bar y$, we set $C_1 = i + 2\xi/\pi$ with $\xi \in \Re$, which makes $\bar\sigma(\bar y)$ as an apparently real form:
\begin{equation}
\label{varsigma:3}
\bar \varsigma(\bar y) \equiv  \frac{1}{\bar y}+  	\frac{\xi \left[ I_0(\bar y) +  I_2(\bar y) \right]
		- \left[ K_0(\bar y) + K_{2} ( \bar y) \right]	 }{ \xi I_1( \bar y ) + K_1(\bar y) }  \,.
\end{equation}
We note that Eq.~\eqref{varsigma:3} satisfies the required differential equation, $- 2\bar\varsigma'- \bar \varsigma^2 = -4- 3/ \bar y^2 $, where $'$ denotes the differentiation with respect to $\bar y$.
Integrating the $\bar \varsigma$ function give
$$
\int \varsigma dy = -\int \bar \varsigma(\bar y) d\bar y
= - \log \bar y - 2\log (\xi I_1(\bar y) + K_1(\bar y)) \,.
$$
Therefore, we get the $\varSigma$ function from the second equation in Eq.~\eqref{Riccati2},
\begin{equation}
\label{varSigma:3}
\varSigma(y) = \frac{\varSigma_0 e^{2\bar y}}{\bar y}
		\frac{1}{\big(\xi I_1(\bar y) +K_1(\bar y)\big)^2} \,.
\end{equation}
For $\bar y \ll 1$, the two Bessel functions behaves as
$$
K_1(\bar y) \approx \frac1{\bar y} +\frac{-1+ 2\gamma +2\log\frac{\bar y}2}{4} \bar y + O(\bar y^3),
\qquad
I_1(\bar y) \approx \frac{\bar y}{2} + O(\bar y^3) \,,
$$
where $\gamma$ denotes the Euler-gamma.
Therefore, in the $G_0 \sim 0$ limit, unless $\xi \to \infty$, the $\varSigma$ functions behave as
\begin{equation}
\label{varSigma:31}
\varSigma(y) \approx
 \frac{G_0\varSigma_0}{4} y \left(1+ \frac{G_0 y}{2} + O(\bar y^2) \right) \,.
\end{equation}

\end{enumerate}

\begin{enumerate}

\item[(b)] On the other hand, when
\begin{equation}
\label{G0=0}
G_0 =0 \,,
\end{equation}
the solution to Eq.~\eqref{Sigma:eq1-1} with the second equation in Eq.~\eqref{Riccati2} has a closed form, with $g_0 =1$ as in Eq.~\eqref{g0:iii},
\begin{equation}
\label{Sigma:sol1-11}
\varSigma = \frac{\varSigma_0 y}{(y_0 y^2+1)^2} \,,
\end{equation}
where $y_0$ is an integration constant. Let us describe the simplest forms:
\begin{enumerate}
\item The typical choice $\varSigma =y$ appears and when we choose $y_0 = 0$ additionally, where the constant $\varSigma_0$ is absorbed into the arbitrariness of $\sigma$ and $Q$.

\item The next simplest choice is the case $\varSigma_0 =\varSigma_0' y_0^2 \to \infty$ which leads $\varSigma = \varSigma_0' y^{-3}$.
\end{enumerate}

The first case corresponds to the limiting case in Eq.~\eqref{varSigma:31} and the second to the limit $G_0 \to 0$ after taking $\xi \to \infty$.

Notice that when $y_0 > 0$, $\varSigma$ has a maximum value $9\varSigma_0/(16\sqrt{3y_0})$ at $y= 1/\sqrt{3y_0}$.
Since $g_{\theta\theta} = \Sigma = P \varSigma$ plays the role of a circumferential radius, the geometry described by the metric with this choice of $\varSigma$ does not admit an asymptotic region.
When $y_0< 0$, the asymptotic regions appear as $y \to \pm (-y_0)^{-1/2}$.

\end{enumerate}

\begin{enumerate}

\item[iv)] \label{Sigma:case4} When $g_{k\geq 3}=0$, the equation becomes
$$
\frac23 \mathring \varsigma -\frac13 \varsigma^2 = - \frac{g_0}{y^2}+  g_2- \frac{G_0^2}{4} \,.
$$
The solution is described by Bessel functions when $g_2 \neq G_0^2/4$ similarly as the case~iii).
The same solutions happens as in Eq.~\eqref{varsigma:2-a}.
As shown in Case~iv)  in Sec.~\ref{sec:4}, there is no consistent solution for $\Gamma$ unless $\dot p=0$.
Since the case $\dot p = 0$ is treated separately, we disregard the possibility $g_{k \geq 3} = 0$ here.

\end{enumerate}

\begin{enumerate}
\item[v)] When $\dot p =0$:\\
As discussed in the previous section, this case requires special care. Here, $g_{k \neq 1} = 0$ for all $k \geq 0$, which coincides with the condition of Case (ii). Consequently, the solution for $\varSigma$ coincides with that obtained in Case (ii).
\end{enumerate}

\begin{enumerate}

\item[vi)] \label{Sigma:case5}
Even though, we do not have a general method in solving the Riccati equation~\eqref{Riccati2}, given a specific solution $\varsigma_1(y)$, we can construct a general solution by using the following technique. First, we change variable from $\varsigma$ to $u$:
\begin{equation}
\label{varsigma:gen}
 \varsigma = \varsigma_1(y)+ \frac1{u(y)}\,.
\end{equation}
Then, we get the linear differential equation, $2\mathring u + 2\varsigma_1 u + 1=0 $,  for $u$.
This equation is integrable in general and its solution is
\begin{equation}
\label{u}
u = e^{-\int_{y_0}^y\varsigma_1(y') dy'} \left[u_1 -\frac12 \int_{y_0} ^ye^{\int_{y_0}^{y''}\varsigma_1(y') dy'} dy''\right] \,,
\end{equation}
where $u_1$ is an integration constant and $y_0$ is an arbitrary number.
Therefore, one can try various trial solutions starting from any specific solution $\varsigma_1(y)$ by setting the right-hand side of Eq.~\eqref{Riccati2} to $\frac23 \mathring \varsigma_1- \frac13 \varsigma_1^2$.

\vspace{.2cm}
To illustrate this method, we pose two specific solutions and develop the corresponding general solutions.

\begin{enumerate}

\item[(a)]
Explicitly, let us choose $\varSigma =y$ as a specific solution.
Then, we can easily find a specific solution $\varsigma_1 = 1/y$ and
$$
u(y) = \frac{\alpha}{y} -\frac{y}{4} \,.
$$
The resulting general solution for $\varSigma$ is nothing but that in Eq.~\eqref{Sigma:sol1-11}.

\end{enumerate}

\vspace{.2cm}
\begin{enumerate}
\item[(b)]
We can also apply this method to the specific case of the solutions~\eqref{varsigma:3} to find a general solutions..
When $\xi =0$, the function $\varSigma$ becomes
\begin{equation}
\label{varSigma4}
\bar \varsigma_{\xi=0}(\bar y) \equiv  \frac{1}{\bar y}- \frac{  K_0(\bar y) + K_{2} ( \bar y)}{  K_1(\bar y)} \,.
\end{equation}
Then, from Eq.~\eqref{u}, we get
$$
u=  \bar y K_1(\bar y)^2 \left[u_1 - \frac{2}{G_0} \frac{I_1(\bar y)}{K_1(\bar y)} \right] \,.
$$
The general form for $\varsigma $ becomes
\begin{equation}
\label{varsigma11}
\varsigma =
	-\frac{G_0}{4} \left( \frac1{\bar y} - \frac{K_0(\bar y)+ K_2(\bar y)}{K_1(\bar y)}\right)
	+ \frac{1}{  \bar y K_1(\bar y)\left[u_1 K_1(\bar y) - \frac{2}{G_0} I_1(\bar y) \right] } \,,
\end{equation}
Finally, the function $\varSigma$ becomes
\begin{equation}
\label{varSigma11}
\varSigma (y) =\frac{4\varSigma_0 e^{\frac{G_0 y}{2}}}{G_0 y}
	\frac{1}{K_1(\bar y)^2} \exp\left[\frac{4}{G_0} \int ^{\bar y} \frac{1}{\bar y K_1(\bar y) [u_1 K_1(\bar y) - \frac{2}{G_0} I_1 (\bar y)]} d\bar y \right] \,.
\end{equation}
Integrating the exponents, we get
\begin{equation}
\label{varSigma12}
\varSigma (y)
=\frac{4\varSigma_0 e^{\frac{G_0 y}{2}}}{G_0 y}
	\frac{1}{
		\left[ u_1 K_1(\bar y)
		-\frac{2}{G_0} I_1(\bar y)\right]^2}   \,,
\end{equation}
which is identical to Eq.~\eqref{varSigma:3}.
As shown in these example, one can develop various solutions starting from a specific solution.

\end{enumerate}

\end{enumerate}

\end{enumerate}
Solving the equation~\eqref{Riccati2} with the $\Gamma$ equation in Eq.~\eqref{app:Gamma:gen}, we obtain the solution that
allows for  the separability structure.

\section{An example:  Generalized Kerr-Newman-Nut solution in black hole/wormhole combination } \label{Sec6}

\quad
As an explicit illustration of the formalism developed in the previous sections, we construct a generalized Kerr-Newman-NUT-type solution describing a black hole-wormhole composite geometry.
This example serves as a nontrivial consistency check of our framework: it preserves the stationary and axisymmetric structure characteristic of Kerr-Newman-NUT spacetimes, while allowing for nontrivial topology and the emergence of a throat connecting two asymptotic regions.

\vspace{.3cm}
The class of geometries permitted by Eqs.~\eqref{metric:gen} and \eqref{R12=01}  is broad.
These equations constrain only part of the functional forms of $P(x)$, $\varSigma(y)$, $q(x)$, $Q(x)$, $\sigma(r)$, and $p(x)$, while leaving $\Delta(r)$ unconstrained.
Consequently, the geometry is not uniquely fixed.

To obtain physically relevant configurations, we impose additional constraints motivated by the matter sector.
Rather than specifying a particular matter a priori, we constrain the Einstein tensor (and hence the stress-energy tensor) directly.
Specifically, we first require that the Einstein tensor be diagonal in the orthonormal frame~\eqref{basis}.
We then impose either angular isotropy, $G_{\hat 2\hat 2}=G_{\hat 3\hat 3}$, or the radial equation of state $w_1 \equiv p_r/\rho = -1$.
Although these conditions do not fully determine the geometry, they significantly restrict the function $\sigma(r)$ and yield nontrivial relations among the angular functions.

This solution appears when $\varSigma =y$ is given by the case~iii) in Eq.~\eqref{Sigma:sol1-11} with $y_0=0$ and $\Gamma$
by the case iii)-(a) in Eq.~\eqref{Gamma:case1-1}. The function $p(x)$ is given in Eq.~\eqref{p:case1-1}.
The value $G_0 =0$ as in Eq.~\eqref{G0=0} which leads $q=P$. Therefore, the metric functions satisfy
\begin{equation}
\label{ansatz1}
\varSigma = y, \quad \Gamma = a^2c_1\left(c_2+\sigma\right), \quad p = c_1 \left(c_2-Q\right), \quad
q = P, \quad \Gamma - a^2 p = a^2c_1 y \,,
\end{equation}
where $P(x)$, $Q(x)$, and $\sigma(r)$ are free functions.
Putting these results to Eq.~\eqref{metric:gen} and removing the irrelevant constants $c_1$ and $c_2$ by performing the coordinate transform,
$ \frac{t}{ac_1} - (c_2-a^2) \phi \to a t $, we get the metric
\begin{equation}
\label{metric:case1-2}
ds^2 = -  \frac{ \Delta}{  y } 	\left( dt -  \frac{(a^2-Q)}{a} d\phi \right)^2 + \frac{y}{\Delta }dr^2 + P y d\theta^2
 + \frac{P \sin^2\theta }{ y} \left(adt - (a^2+\sigma) d\phi \right)^2 \,.
\end{equation}
Here, $y(r,x)= \sigma(r) + Q(x)$, $P=P(x)$ and $\Delta \equiv \Delta(r)$.

The Einstein tensor component $G_{\hat 1 \hat 2} =0$ as expected for this metric.
However, there is a non-vanishing off-diagonal component of the Einstein tensor,
\begin{equation}
\label{G03:case1-1}
G_{\hat 0\hat 3} = \frac{\sin (\theta ) \sqrt{P \Delta } }{2 a
   (Q+\sigma )^2}
   \left(a^2 \sigma''(r)+ \frac{\dot Q}{P^2} \frac{d}{dx} \log \frac{P }{\dot Q}\right) \,.
\end{equation}
This result definitely indicates that if we want the stress tensor to have a diagonal form with respect to the orthonormal basis, it requires
\begin{equation}
a^2 \sigma''(r)+ \frac{\dot Q(x)}{P^2(x)} \frac{d}{dx} \log \frac{P(x) }{\dot Q(x)}=0 \,. \nonumber
\end{equation}
Noting $\sigma(r)$ is a function of $r$ only and $P$, $Q$ are functions of $x$ only, we easily get
\begin{equation}
\label{G03=0}
\sigma'' = 2\sigma_2, \qquad
\ddot Q= \frac{\dot P}{P}\dot Q+ 2a^2 \sigma_2 P^2 \,,
\end{equation}
where $\sigma_2$ is a real constant introduced for separation of variables. When $\dot P= 0$, this equation determines the function $ Q(x)$.
Notice that the first equation determines the function $\sigma(r)$ to be at most quadratic to $r$.
Let us set $\sigma = \sigma_2 (r-r_0)^2 + \sigma_0$. Notice that we may always set $r_0 =0$ by using the change of variables $r-r_0 \to r$ because the functional dependence of $\Delta(r)$ is not fixed yet. We may also set $\sigma_0 = 0$ because the function $Q$ is not identified and we can always scale $a$. Assuming $\sigma_2 > 0$, we absorb $\sigma_2$ by setting $\sqrt{\sigma_2} r \to r$.
Therefore, we set $y = r^2 + Q(x)$ with $\sigma = r^2$ without loss of generality.
The metric with diagonal stress tensor now takes the form,
\begin{eqnarray}
\label{metric:case2-3}
ds^2 = &&  -  \frac{ \Delta}{  r^2+Q } \left( dt -  \frac{(a^2-Q)}{a} d\phi \right)^2 	+ \frac{r^2+Q}{\Delta }dr^2
+ P (r^2+Q) d\theta^2  \nonumber \\
      && + \frac{P \sin^2\theta }{ r^2+Q} \left(a dt - (r^2+a^2) d\phi \right)^2 \,.
\end{eqnarray}

The Einstein tensor for this metric is still complex to write down explicitly.
However, the following combinations are simple:
\begin{equation}
G_{\hat 0\hat 0}+ G_{\hat 1\hat 1} = \frac{\Delta (\dot Q^2- 4a^2 P^2Q)}{2a^2 P^2 (r^2 +Q)^2}\,, \qquad
G_{\hat 2\hat 2} - G_{\hat 3\hat 3} = \frac{(1-x^2) ( \dot Q^2- 4a^2 P^2 Q )}{2 P (r^2 + Q)^3} \,. \nonumber
\end{equation}
Noting the forms of the two combinations, imposing angular isotropy, $G_{\hat 2 \hat 2} = G_{\hat 3 \hat 3}$, also yields $w_1 \equiv p_r/\rho = -1$. Requiring angular isotropy,
\begin{equation}
\label{P:dQ}
P^2 = \frac{(\dot {\sqrt{Q}})^2}{a^2 }  \quad \rightarrow \quad
\frac{d {\sqrt{Q}}}{dx} = \pm a P \,,
\end{equation}
and substituting this relation into the second equation of Eq.~\eqref{G03=0}, we find that the latter is automatically satisfied.

\vspace{.2cm}
Given $P(x)$, one can determine $Q(x)$ from Eq.~\eqref{P:dQ}. Then, the metric is fully fixed when $\Delta(r)$ is given.
When $P=1$, Equation~\eqref{P:dQ} gives $Q(x) =a^2 ( \cos \theta + q_0)^2 $, in which $q_0$ is an integration constant.
If we choose $q_0=0$ and $\Delta = r^2 +a^2- 2M r$, the metric describes the Kerr geometry.
Starting from this known geometry, we show how far the metric~\eqref{metric:case2-3} can be generalized.
We first let $\Delta= r^2+ a^2-2Mr + v(r)$, where $v(r)$ is a free function of $r$.
Then, the Einstein tensor becomes
\begin{equation}
G_{\hat 0\hat 0} =-G_{\hat 1\hat 1}= \frac{v- rv'}{y^2} \,, \qquad
G_{\hat 2\hat 2} = G_{\hat 3\hat 3}= \frac{v - rv ' +\frac12 (r^2+ a^2 x^2) v'' }{y^2} \,. \nonumber
\end{equation}
The properties for this case is well established in Ref.~\cite{Kim:2025sdj} including various matters including electro-magnetic field.
Therefore, we consider other generalizations.
The constant $q_0 \neq 0$ plays the role of a nut charge and the metric~\eqref{metric:case2-3} describes the Kerr black hole having nut charge.
When $\Delta_0 = a^2(1-q_0^2)$, the typical vacuum Kerr-Nut solution appears.

 \vspace{.3cm}
The Einstein tensor of the general form of the metric~\eqref{metric:case2-3} is
\begin{eqnarray}
\label{Gkk}
G_{\hat 0\hat 0} &=& \frac{1}{y^3}\left\{\left[3 \Delta + a^2(x^2-1)P\right] \left(\frac{ \dot Q^2}{4 a^2 P^2}- Q\right)
                 + \frac{(r^2+Q)^2}{P} 	\left[\frac{x \dot P}{P}+1 		+\frac{1}{2} \left(x^2-1\right)
   		\left(\frac{\ddot P}{P}-\frac{ \dot P^2}{P^2}\right) 	   \right] \right. \nonumber \\
      &&\left. 	   + \left(Q+r^2\right) \left[ \Delta -r \Delta'	+a^2(x^2-1)  P 	- \frac{Q }{P}\left(
				\frac{x^2-1}{2} 	\frac{ \dot P \dot Q}{ PQ}	+ \frac{x \dot Q}{Q}\right)	\right] \right\} \,, \nonumber \\
G_{\hat 2\hat 2}&=& \frac1{y^3}  \left\{ 	\left[\Delta+ a^2(x^2-1)P  \right] \left(\frac{\dot Q^2}{4 a^2P^2}-Q\right)
  		+ \left(Q+r^2\right)^2 \frac{ \Delta''}{2} \right. \nonumber \\
&& \left. +(Q+r^2)\left[ \Delta-r \Delta' 	+a^2\left(x^2-1\right)  P  -\frac{Q}{P}\left(\frac{x^2-1}2\frac{\dot P \dot Q}{P Q}
	 	+\frac{x \dot Q}{Q}\right) \right] \right\} \,.
\end{eqnarray}
In the following subsections, we discuss how to determine the metric including the functions $P(x)$ and $\Delta(x)$ starting from the Einstein tensor.

\subsection{Angular isotropic black holes with electromagnetic field-like matter}

\quad
In this subsection, we construct metric for the matter stress tensor satisfies $w_1= -1$ and $w_2=w_3 =1$, which is that of the electromagnetic field.
We do not solve the electromagnetic field equation but concentrate on the Einstein tensor~\eqref{Gkk} itself.

Notice that the second lines of the two components of the Einstein tensor in Eq.~\eqref{Gkk} are the same but the first lines are different.
When we require the angular isotropy, the terms containing  $ \frac{\dot Q^2}{4a^2P^2}-Q$ vanishes and get
\begin{eqnarray}
\label{G:bh}
G_{\hat 0\hat 0} &=& \frac{1}{y^2}\left\{ 	\frac{y}{P} \left[\frac{x \dot P}{P}+1 	+\frac{1}{2} \left(x^2-1\right)
   		\left(\frac{\ddot P}{P}-\frac{ \dot P^2}{P^2}\right)  \right] 	+ \Delta -r \Delta'
	 +a^2(x^2-1)  P\left(1	- \frac{ \dot P \dot Q}{ 2a^2 P^3}\right) -  \frac{x \dot Q}{P}	 \right\} \,, \nonumber \\
G_{\hat 2\hat 2}&=& \frac1{y^2}  \left\{\frac{1}{2} y \Delta'' 	+ \Delta-r \Delta '
  +a^2\left(x^2-1\right)  P\left(1 	-\frac{\dot P \dot Q}{2 a^2 P^3 }\right) -\frac{x \dot Q}{P} \right\} \,.
\end{eqnarray}
Comparing the two components, one may easily notice that $G_{\hat 0\hat 0} = G_{\hat 2\hat 2}$ when
\begin{equation}
\label{w2=1}
\frac{1}{P} \left[\frac{x \dot P}{P}+1 		+\frac{1}{2} \left(x^2-1\right)
   		\left(\frac{\ddot P}{P}-\frac{ \dot P^2}{P^2}\right)    \right] = \frac12\Delta''(r) \,.
\end{equation}
Notice that the left/right-hand side is a function of $x/r$ only, respectively.
Therefore, this equation holds only when $\Delta''$ is independent of $r$ making $\Delta(r)$ to be a quadratic function of $r$:
\begin{equation}
\label{w2=1 1}
\Delta''(r) = 2\Delta_2 \quad \rightarrow \quad \Delta = \Delta_2 (r^2 -2M r + \Delta_0) \,.
\end{equation}
Given this result, the angular part equation, by introducing
\begin{equation}
\label{z cP}
\partial_z \equiv (x^2-1) \partial_x \,, \qquad \c P \equiv (1-x^2) P \,,
\end{equation}
takes the following simpler form:
\begin{equation}
\label{angular eq}
\frac{d^2 \log \c P}{dz^2}= -2\Delta_2 \c P \,.
\end{equation}
Before going further, we examine the angular part of the metric~\eqref{metric:case2-3}.
Let us introduce a proper angular coordinate $\chi $ satisfying $d\chi \equiv \sqrt{P} d\theta$.
Then, the angular part of the metric can be written as
$$
P\left[ d\theta^2  + \frac{ \sin^2\theta }{ (r^2+Q)^2} 	\left(a dt - (r^2+a^2) d\phi \right)^2 \right]
=d\chi^2  + \frac{ \c P(\chi) }{ (r^2+Q)^2} 	\left(a dt - (r^2+a^2) d\phi \right)^2  \,.
$$
Thus, the function $\c P(\chi)$ as a function of angle variable $\chi$ explicitly represents the angular geometry.
In Appendix~\ref{App1}, we solve this differential equation~\eqref{angular eq} in a general form and find the function $\c P(\chi)$ with $\beta = 2\Delta_2$:
\begin{eqnarray}
\label{P}
\c P(\chi) \equiv \alpha^2
\left\{
\begin{aligned}
 &\sin^2 \left[\sqrt{\Delta_2}\chi\right]   &\qquad \Delta_2 >  0
 \\
 & \quad \chi^2  & \qquad  \Delta_2 = 0
 \\
 & \sinh^2 \left[\sqrt{|\Delta_2|} \chi\right]   & \qquad  \Delta_2 < 0
 \end{aligned}
 \right. \,,
\end{eqnarray}
where the symmetric axis is chosen to be located at $\chi=0$ and $\alpha$ is an integration constant, precisely specified in App.~\ref{App1}. 
Notice that each case corresponds to the open, flat, and the closed 2-dimensional geometry, respectively.
Finally, the $Q$ function from Eq.~\eqref{P:dQ} becomes, noting $dw/dz = -\beta \c P$,
\begin{eqnarray}
\label{Q:3}
Q = a^2 \left(\int \c P dz\right)^2
= \frac{a^2\alpha^2}{\Delta_2}\left(q_0 -w \right)^2  ; \qquad
 w= \left\{
 		\begin{aligned}
		&\cos \left[\sqrt{\Delta_2} \chi\right]\\
		 &\chi^2   \\
		&\cosh \left[\sqrt{\Delta_2} \chi\right]
		 \end{aligned}
 \right. \,,
\end{eqnarray}
where $q_0$ is an integration constant.
We further perform successive change of variables, first $\sqrt{\Delta_2} \chi \to \chi$,  $r \to \sqrt{\Delta_2} r$ and then,
$M/\sqrt{\Delta_2} \to M$, $a/\sqrt{\Delta_2} \to a$, $\Delta_2 \phi \to \phi$, $\sqrt{\Delta_2} t \to t$,  $\Delta_0/\Delta_2 \to \Delta_0$, $\alpha/\Delta_2 \to \alpha$, and $Q/\Delta_2 \to Q$, to get the metric
\begin{eqnarray}
\label{metric:case2-5}
ds^2 &=&  -  \frac{ r^2-2M r+\Delta_0}{  r^2+Q } \left( dt -  \frac{(a^2-Q) }{a}d\phi \right)^2
	+ \frac{  r^2+ Q }{ r^2-2M r+ \Delta_0} dr^2 \nonumber \\
&& + (r^2+ Q)\left[ d\chi^2  + \frac{\c P(\chi) }{  ( r^2+Q)^2}
	\left( a dt - (r^2+a^2) d\phi \right)^2  \right]  \,,
\end{eqnarray}
where $Q = a^2 \alpha^2  (q_0 -w(\chi))^2$ and identification of angular coordinate $\phi$ should be taken into account.
Notice also that the constant $\Delta_2=1$ in this metric at the present form.
When $\alpha=1$, the angular deficit disappears making all the metric form to be identical to that of Kerr-Nut geometry with $q_0$ plays the role of a nut charge.
For closed spacetime with positive $\Delta_2$, the Nut geometry has closed time-like curve to avoid a Misner-string singularity~\cite{Astorino:2023elf}.
When $\alpha \neq 1$ with $q_0=0$, it denotes the presence of a angular deficit along the poles, which describes black holes stretched by tensions due to cosmic string attached to the poles.

The energy density for the metric is
\begin{eqnarray}
\label{rho}
\rho = G_{\hat 0\hat 0}= \frac{a^2\alpha^2 (q_0^2-1) 	+ \Delta_0}{y^2} \,.
\end{eqnarray}
When $\Delta_0 = (1-q_0^2) (a\alpha)^2$, the energy density vanishes and the metric~\eqref{metric:case2-5} describes the vacuum rotating black hole solution having nut-charge and angle-deficit in various $2$-dimensional topology.

\subsection{Rotating black hole solutions with a cosmological constant-like term}

\quad

In this subsection, we construct a metric for a matter stress tensor that satisfies $w_k = -1$ for all $k = 1,2,3$, analogous to a cosmological-constant-like term, but distinct from the well-known Kerr-de Sitter spacetime~\cite{Carter:2009nex, Akcay:2010vt}.
It would be a general solution including the solution shown in Ref.~\cite{Beltracchi:2021ris}, which would correspond to the case of $w = -1$ in Eq.~$(11)$ in Ref.~\cite{Kim:2019hfp}.
For more general ones, refer to Refs.~\cite{Griffiths:2005qp, Ovcharenko:2025fxg}.

Because $w_1 = -1$ and $w_2=w_3$ from Eq.~\eqref{P:dQ}, requiring the condition $G_{\hat 0\hat 0}+ G_{\hat 2\hat2}=0$ is enough, which gives, from Eq.~\eqref{G:bh} and using $y = r^2 + Q$,
\begin{eqnarray}
\label{CC eom}
&&\frac{1}{2} r^2 \Delta'' +2\left[ \Delta -r \Delta '\right] 	+\frac{1}{2} Q \Delta''  +\frac{r^2}{P}
	\left[\frac{x \dot P}{P}+1 		+\frac{1}{2} \left(x^2-1\right)    		\left(\frac{\ddot P}{P}-\frac{ \dot P^2}{P^2}\right)
	   \right] \nonumber \\
&&  +\frac{Q}{P} 	\left[\frac{x \dot P}{P}+1 		+\frac{1}{2} \left(x^2-1\right)
   		\left(\frac{\ddot P}{P}-\frac{ \dot P^2}{P^2}\right) 	   \right]   +2a^2(x^2-1)  P\left(1
 			- \frac{ \dot P \dot Q}{ 2a^2 P^3}\right)    			-  \frac{2x \dot Q}{P}  =0  \,.
\end{eqnarray}
Noting the $r$-independent terms and the $r^2$-proportional terms, this equation constrain the second derivative $\Delta''$ cannot have higher order terms than $r^2$ and $O(r)$ term to vanish, i.e.,
\begin{equation}
\label{CC Delta}
\Delta = \frac{r^4}{\ell^2} + \Delta_2 r^2 + \Delta_1 r + \Delta_0 \,.
\end{equation}
Putting this result to the equation~\eqref{CC eom}, we get, in series of $r^2$,
\begin{eqnarray}
\label{eomL1}
&&  \frac{r^2}{P} 	\left[\frac{x \dot P}{P}+1 		+\frac{1}{2} \left(x^2-1\right)    		\left(\frac{\ddot P}{P}-\frac{ \dot P^2}{P^2}\right)
	-\Delta_2 P+ \frac{6}{\ell^2} PQ 	\right] \nonumber \\
&&  +\frac{Q}{P} 	\left[\frac{x \dot P}{P}+1 		+\frac{1}{2} \left(x^2-1\right)
   		\left(\frac{\ddot P}{P}-\frac{ \dot P^2}{P^2}\right) 	   \right]
  +2a^2(x^2-1)  P\left(1  	- \frac{ \dot P \dot Q}{ 2a^2 P^3}\right) 	\nonumber \\
&&  -  \frac{2x \dot Q}{P}	+2\Delta_0 + Q\Delta_2 =0 \,.
\end{eqnarray}
When these equations have solution, the coefficients of each power of $r^2$ should vanish separately:
\begin{eqnarray}
\label{1eq3}
&& \frac{x \dot P}{P}+1 	+\frac{1}{2} \left(x^2-1\right)
   		\left(\frac{\ddot P}{P}-\frac{ \dot P^2}{P^2}\right) =\left(\Delta_2 - \frac{6}{\ell^2} Q\right)P ,  \\
&& \frac{Q}{P} 	\left[\frac{x \dot P}{P}+1 		+\frac{1}{2} \left(x^2-1\right)
   		\left(\frac{\ddot P}{P}-\frac{ \dot P^2}{P^2}\right)   \right]
  +2a^2(x^2-1)  P\left(1 	- \frac{ \dot P \dot Q}{ 2a^2 P^3}\right)
   			-  \frac{2x \dot Q}{P}	+2\Delta_0 + Q\Delta_2	  =0 \,. \nonumber
\end{eqnarray}
Putting the first equation~\eqref{1eq3} to the second, we get
\begin{eqnarray}
\label{1eq1}
a^2(x^2-1)  P\left(1 - \frac{ \dot P \dot Q}{ 2a^2 P^3}\right)    	-  \frac{x \dot Q}{P}
	+Q \left[\Delta_2 - \frac{3}{\ell^2} Q\right] 	+\Delta_0 =0 \,.
\end{eqnarray}
In addition, these two equations should be compatible with the angular isotropic condition in Eq.~\eqref{P:dQ}.
The three equations are complex at the present form.
However, when we introduce variable $z$ and the function $\c P(z)$ as in Eq.~\eqref{z cP},
the three equations take neater forms:
\begin{eqnarray}
\label{angular eq2}
&&\frac12 \frac{d^2 \log \c P}{dz^2}= -\left(\Delta_2 - \frac{6}{\ell^2} Q\right) \c P \,, \nonumber \\
&&  \frac{\partial_z \c P \partial_z Q}{\c P^2} +2Q \left[\Delta_2 - \frac{3}{\ell^2} Q\right]
  -2a^2\c P +2\Delta_0 =0 \,,   \nonumber \\
&& \c P^2 = \frac{(\partial_zQ)^2}{4a^2 Q} \,.
\end{eqnarray}
In deriving the second equation, we have used
\begin{eqnarray}
(x^2-1)  P\left(1 - \frac{ \dot P \dot Q}{ 2a^2 P^3}\right)
&=&-\c  P +\frac{\partial_z \c P \partial_z Q}{2a^2 \c P^2}
	- \frac{2x \partial_z Q}{2a^2 \c P} \,.
\end{eqnarray}
Now, the two functions $\c P$ and $Q$ satisfies three equations given in Eq.~\eqref{angular eq2}.
Fortunately, in App.~\ref{App A}, we show that the second equation is a first integral (i.e., a conserved quantity) of the system formed by  the other two, with $\Delta_0$ playing the role of the integration constant determined by the initial conditions.
There, we also solve the coupled differential equation to obtain the solution:
\begin{equation}
\label{pQ}
\frac{dQ}{dz} =-\frac{2\varepsilon}{a} 	\left[\Delta_2Q^{3/2}  - \frac{1}{\ell^2} Q^{5/2} -\Delta_0 Q^{1/2}\right]+ C_1 Q \,.
\end{equation}
Now, we set $\varphi = \sqrt{Q}$. Then, the equation becomes
\begin{equation}
\frac{d\varphi}{dz} =-\frac{\varepsilon}{a} 	\left[\Delta_2 \varphi^{2}  - \frac{1}{\ell^2} \varphi^{4}
		-\Delta_0  \right]+ \frac{C_1}{2} \varphi \,.   \nonumber
\end{equation}
Therefore, the right-hand side is a quartic polynomial in $\varphi$.
The function $\c P$ becomes
\begin{equation}
\label{P:B}
\c P = \frac{\varepsilon Q_{,z}}{2a \sqrt{Q}} =\frac{\varepsilon}{a} \frac{d\varphi}{dz}
=  -\frac{1}{a^2} 	\left[\Delta_2 \varphi^{2}  - \frac{1}{\ell^2} \varphi^{4}
		-\Delta_0  \right]+ \frac{\varepsilon C_1}{2a} \varphi \,.
\end{equation}
Putting these results to the metric~\eqref{metric:case2-3}, we get
\begin{eqnarray}
\label{metric:case2-8}
ds^2 &=&  -  \frac{ \frac{r^4}{\ell^2} + \Delta_2 r^2 + \Delta_1 r + \Delta_0}{  r^2+\varphi^2}
	\left( dt -  \frac{(a^2-\varphi^2)}{a} d\phi \right)^2 	+ \frac{r^2+\varphi^2}{\frac{r^4}{\ell^2} + \Delta_2 r^2 + \Delta_1 r + \Delta_0 }dr^2 \nonumber \\
&& + \frac{r^2+\varphi^2}{\frac{\varphi^4}{\ell^2} 	-\Delta_2\varphi^2+\Delta_0 +\frac{\varepsilon aC_1}{2} \varphi } d\varphi^2
   + \frac{\frac{\varphi^4}{\ell^2}  -\Delta_2 \varphi^{2} 	+\Delta_0 + \frac{\varepsilon aC_1}{2} \varphi }{ a^2( r^2+\varphi^2)}
	\left(a dt - (r^2+a^2) d\phi \right)^2  \,.
\end{eqnarray}
Here, we use $P d\theta^2 = \c P dz^2 = \c P\left(\frac{dz}{d\varphi}\right)^2 d\varphi^2= \frac{\varepsilon}{a} \frac{dz}{d\varphi} d\varphi^2$.

\subsection{Rotating black hole with anisotropic matters and global monopole}

\quad

In this subsection, we consider the black hole geometry having global monopole which consists of two anisotropic matters,
one is electromagnetic field-like and the other has equation of state  $w_2 \neq 1$.

We note that the $1/y^2$ terms of the Einstein tensor components $G_{\hat 0\hat0}$ and $G_{\hat 2\hat2}$ in Eq.~\eqref{G:bh} are identical.
Therefore, the matter responsible for this contribution must satisfy $w_2 = 1$ with $w_1= -1$.
On the other hand, the $1/y$ terms in the Einstein tensor components may originate from a different source with a different equation-of-state parameter $w_2$.

From the Einstein tensor~\eqref{G:bh}, we write the $1/y$ parts of $G_{\hat 0\hat 0}$ and $G_{\hat 2\hat 2}$ as
\begin{equation}
\rho_2 \equiv \frac{1}y \mbox{ part of } G_{\hat 0\hat 0} =  \frac{1}{Py}
	\left[\frac{x \dot P}{P}+1
		+\frac{1}{2} \left(x^2-1\right)
   		\left(\frac{\ddot P}{P}-\frac{ \dot P^2}{P^2}\right)
	   \right], \qquad p_2 \equiv \frac{1}y \mbox{ part of } G_{\hat 2\hat 2}
	= \frac{\Delta''}{2y} \,,  \nonumber
\end{equation}
where $\rho_2/p_2$ represent the energy density/pressure along $\hat \theta$ direction of the second matter, not the total ones.
Now, we require the equation of state for the second matter for the angular direction to be $w_2=$ constant, which may not be one.
This condition presents an equation
\begin{equation}
\frac{w_2}{P} 	\left[\frac{x \dot P}{P}+1 		+\frac{1}{2} \left(x^2-1\right)
   		\left(\frac{\ddot P}{P}-\frac{ \dot P^2}{P^2}\right)   \right] = \frac{\Delta''}{2} \,. \nonumber
\end{equation}
Solving this equation, $P$ and $Q$ are given by the equations~\eqref{P} and \eqref{Q:3} with appropriate  $w_2$ scalings and $\Delta = \Delta_2(r^2-2Mr +\Delta_0) $.

The metric with diagonal stress tensor now takes the form,
\begin{eqnarray}
\label{metric:bh2-1}
ds^2 &=&  -  \frac{ \Delta}{  r^2+Q } 	\left( dt -  \frac{(a^2-Q)}{a} d\phi \right)^2
	+ \frac{r^2+Q}{\Delta }dr^2  \nonumber \\
    &&    +\frac{w_2(r^2+Q)}{\Delta_2} \left[ d\chi^2
 + \frac{\c P(\chi) }{( r^2+Q)^2} 	\left( a dt - (r^2+a^2) d\phi \right)^2 \right]  \,.
\end{eqnarray}
Using the same redefinitions, $r \to \sqrt{\Delta_2} r$ and  additionally change notation $M/\sqrt{\Delta_2} \to M$, $a/\sqrt{\Delta_2} \to a$, $\Delta_2 \phi \to \phi$, $\sqrt{\Delta_2} t \to t$,  $\Delta_0/\Delta_2 \to \Delta_0$, $\alpha/\Delta_2 \to \alpha$, and $Q/\Delta_2 \to Q$ we can set $\Delta_2 =1$ without loss of generality.

Notice that this form of metric is identical to that in Eq.~\eqref{metric:case2-5} except the $w_2$ factor in front of the angular metric, which presents an area rescaling similar to that of the global monopole~\cite{Barriola:1989hx}.
Notice that we cannot scale away this $w_2$ factor.
The presence of $\c P(\chi)$ term denotes that this metric also includes all the topological black holes.

\subsection{Rotating wormholes and angular isotropy breaking}

\quad
In the previous subsections, we have derived various black hole solutions.
Now, we show that the present metric form~\eqref{metric:case2-3} is compatible with
a rotating wormhole geometry at the cost of angular isotropy.

To this end, we first set
\begin{equation}
\label{Q:bQ}
Q = b^2 + \bar Q \,,
\end{equation}
and require $\bar Q$ to satisfy the equation~\eqref{P:dQ} rather than $Q$.
Notice that the absence of the off-diagonal term~\eqref{G03=0} still holds automatically even with this modification.
Remember that Eq.~\eqref{P:dQ} comes from the angular isotropy condition, $w_2 =w_3$.
Therefore, this requirement~\eqref{Q:bQ} breaks the condition.
The components of the Einstein tensor now satisfy
\begin{equation}
\label{wh0}
G_{\hat 0\hat 0}+ G_{\hat 1\hat 1} = \frac{- 2b^2\Delta }{ (r^2 +b^2 + \bar Q)^3}, \qquad
G_{\hat 2\hat 2} - G_{\hat 3\hat 3} = \frac{- 2a^2 b^2 (1-x^2) P  }{  (r^2 +b^2+ \bar Q)^3} \,.
\end{equation}
Notice that the angular isotropy is recovered for a non-rotating system with $a=0$ but the condition $w_1 =-1$ is not.
When $b=0$ (absence of a wormhole throat), both the angular isotropy and $w_1=-1$ are recovered.
As will be shown later, this slight modification develops wormhole throat at $r=0$.

Let us rewrite the Einstein tensor~\eqref{Gkk} of the metric~\eqref{metric:case2-3} with the function $Q$ in Eq.~\eqref{Q:bQ},
\begin{eqnarray}
\label{metric:bh2-1}
G_{\hat 0\hat 0} &=& -\frac{b^2 \left[3 \Delta - a^2\c P\right] }{y^3} + \frac{\Delta -r \Delta'  -a^2 \c P
 		+ \frac{1 }{2\c P} \frac{dQ}{dz}\frac{ d\log \c P}{ dz}}{y^2} -\frac{1 }{2y \c P} \frac{d^2 \log\c P}{dz^2}  \,, \nonumber \\
 G_{\hat 2\hat 2}&=& -\frac{b^2 \left[\Delta- a^2\c P  \right] }{y^3}  + \frac{ \Delta-r \Delta' -a^2\c P
	  +\frac{1 }{2\c P} \frac{dQ}{dz} \frac{ d\log \c P}{ dz} } {y^2}   + \frac{ \Delta''}{2 y}  \,,
\end{eqnarray}
where  the off-diagonal components vanishes and the other two diagonal components can be obtained from Eq.~\eqref{wh0}.
Notice that the first terms of $G_{\hat 0\hat 0} $ and $G_{\hat 2\hat 2}$, which are proportional to $b^2$, develop the wormhole throat.
Therefore, the other terms come from ordinary matters.
Because of this fact, we require that the remaining matters satisfies
\begin{equation}
\bar G_{\hat 0\hat 0 } = \bar G_{\hat 2 \hat 2} \quad \rightarrow \quad
G_{\hat 0\hat 0} +\frac{3b^2  \Delta  }{y^3} =  G_{\hat 2\hat 2}+\frac{b^2 \Delta  }{y^3} \,, \nonumber
\end{equation}
where the barred quantities denote the Einstein tensor derived from the remaining matter.
This requirement leads the angular functions $P(x)$ and $\bar Q(x)$ satisfying the same equation as Eq.~\eqref{w2=1}.
Therefore, the functions $P(x)$ and $\bar Q(x)$ take the same form as in Eqs.~\eqref{P} and \eqref{Q:3} with $Q \to \bar Q$.
After taking the redefinition $r \to \sqrt{\Delta_2}r$ to set $\Delta_2 =1$, we get the metric
\begin{eqnarray}
\label{metric:case2-5wh}
ds^2 &=&  -  \frac{ r^2-2M r+\Delta_0}{  r^2+Q } 	\left( dt -  \frac{(a^2-Q) }{a}d\phi \right)^2
	+ \frac{  r^2+ Q }{ r^2-2M r+ \Delta_0} dr^2 \nonumber \\
&& + (r^2+ Q)\left[ d\chi^2  + \frac{\c P(\chi) }{  ( r^2+Q)^2} 	\left( a dt-  (r^2+a^2) d\phi \right)^2  \right]  \,,
\end{eqnarray}
where
\begin{equation}
\label{Q:wh}
Q(x) = b^2+ a^2\alpha^2 (q_0 - w)^2 \,,
\end{equation}
and $\c P(\chi)$ and $w(\chi)$ are given in Eqs.~\eqref{P} and \eqref{Q:3}, respectively.

To focus on wormhole configurations, we set the NUT charge to zero, $q_0=0$.
After the additional coordinate transformation $t \to t- b^2/a \, \phi$, the metric reduces to
\begin{eqnarray}
\label{metric:wh1}
ds^2 =  -  \frac{ \Delta}{ y} \left( dt -  a\left(1-\alpha^2 w(\chi)^2 \right) d\phi \right)^2
	+ \frac{y}{\Delta }dr^2 + y d\chi^2
	+ \frac{\c P(\chi) }{ y} \left(a dt - (r^2+a^2+ b^2) d\phi \right)^2 \,,
\end{eqnarray}
where
\[
y = r^2 + b^2 + a^2 \alpha^2 w(\chi)^2,
\qquad
\Delta = r^2 - 2Mr + \Delta_0 \,.
\]
The parameter $\alpha \neq 1$ corresponds to the angular deficit mentioned in the black hole solutions, which may be interpreted as a string-like source attached to the wormhole throat.
The function $\c P(\chi)$ encodes the topological structure of the angular sector.

In the static limit ($a=0$), the metric reduces to
\begin{equation}
\label{wh:static}
ds^2 =  -  \frac{ r^2-2M r+ \Delta_0}{  r^2+ b^2}dt^2 	
+ \frac{r^2+b^2}{r^2-2Mr +\Delta_0 }dr^2
+ (r^2+b^2) \left[d\chi^2  + \c P(\chi) d\phi^2 \right] \,,
\end{equation}
which resembles the traversable wormhole model of Simpson and Visser~\cite{Simpson:2018tsi}.

The surface area $A(r)=4\pi(r^2+b^2)$ is minimized at $r=0$, indicating that the wormhole throat is located at this radius.
Depending on the values of $M$ and $\Delta_0$, the geometry may describe either a traversable wormhole or a regular black hole.
In particular, if the horizon lies outside $r=0$, the solution represents a regular black hole, whereas if no horizon forms in the region $r\ge0$, the spacetime corresponds to a wormhole configuration.
Especially, when $M=0$, $\alpha=1$,  and $\Delta_0 = b^2$, the Alice wormhole solution appears.

The nonvanishing components of the Einstein tensor yield
\begin{equation}
8\pi \rho =
\frac{(\Delta_0-2b^2) r^2+ 6b^2Mr +b^2(b^2- 2\Delta_0)}{(r^2+b^2)^3} ,
\qquad
-p_r= p_\theta=p_\phi
= \frac{\Delta_0 r^2+ 2b^2 Mr + b^4}{8\pi (r^2+b^2)^3} \,. \nonumber
\end{equation}
For $M=0$, the energy density cannot remain positive throughout the entire spacetime.
If $\Delta_0 \ge 2b^2$, the asymptotic energy density is nonnegative, but the energy density becomes negative at the throat. Conversely, if $\Delta_0 < 2b^2$, the energy density is negative in the asymptotic region. Thus, the weak energy condition is necessarily violated somewhere in the spacetime, as expected for wormhole geometries.

In summary, the rotating metric~\eqref{metric:wh1} can be interpreted as a rotating generalization of the Alice wormhole solution.

\section{Summary and discussions} \label{Sec7}

\quad
In this work, we investigated the general structure of stationary axisymmetric systems in general relativity, described by a metric of the form $ds^2 = g_{ab}(r,\theta)\, dx^a dx^b.$
A direct treatment of the Einstein equations for such metrics leads to a highly coupled system of nonlinear partial differential equations.
To render the problem tractable, we employed a generalized Carter-type ansatz that facilitates a partial separation of variables.

The resulting metric~\eqref{metric:gen} contains five unknown functions: one function of two variables, $g_{\theta\theta}=\Sigma(r,\theta)$, and four single-variable functions, $\Gamma(r)$, $\Delta(r)$, $q(x)$, and $p(x)$, where $x\equiv\cos\theta$.
We imposed the radial-angular compatibility condition (RACC), namely the vanishing of the mixed component $T_{\hat 1\hat 2}=0$, which implies $G_{\hat 1\hat 2}=R_{\hat 1\hat 2}=0$ in the orthonormal frame~\eqref{basis}.
This condition yields a nonlinear higher-order differential equation involving four functions of $r$ and $\theta$.
Importantly, the function $\Delta(r)$ decouples from this equation.
Rather than solving the resulting partial differential equation directly, we introduced the ansatz $\Sigma(r,x)=P(x)\,\varSigma(y)$ and, $y=\sigma(r)+Q(x),$ which reduces the system to two ordinary differential equations, \eqref{app:dQ:ddq} and \eqref{app:Gamma:gen}, together with a Riccati-type equation~\eqref{Riccati2}.
These reduced equations admit a broad class of solutions; for instance, the equation for $\Gamma$ alone allows six distinct solution branches.

To demonstrate the effectiveness of the method, we constructed an explicit class of solutions corresponding to Eq.~\eqref{ansatz1}.
Requiring the remaining off-diagonal component $G_{\hat 0\hat 3}$ to vanish fixes the radial function to the quadratic form $\sigma(r)=r^2+b^2$ and constrains the angular functions $P(x)$ and $Q(x)$.
Imposing the additional angular isotropy condition $w_2=w_3$, equivalently $w_1=-1$, leads to a rotating black-hole-type geometry.

At this stage, two independent functions remain, $\Delta(r)$ and $P(x)$, which can be determined by specifying an angular equation of state.
For example, choosing $w_2=-1$ corresponds to a cosmological-constant-like contribution, analogous to the case $w=-1$ in Eq.~(11) of Ref.~\cite{Kim:2019hfp}.
More general constructions can be found in Refs.~\cite{Griffiths:2005qp, Ovcharenko:2025fxg}.
Within this framework, we recovered black hole solutions with global monopole-like structure, a Kerr-Newman-type geometry immersed in radiation, and a wormhole solution obtained by relaxing angular isotropy.
For a specific choice of parameters, the latter reduces to the Ellis wormhole.
The solution space of stationary axisymmetric systems is vast, and the present work explores only a restricted sector of it.
Many additional configurations remain to be investigated.

The formalism developed here applies to both vacuum and non-vacuum configurations within the stationary axisymmetric sector.
The reduced field equations incorporate generic matter sources through their effective contributions in Carter's orthonormal frame, allowing for a unified treatment of a broad class of systems.
Owing to its geometric character, the approach may also extend to modified theories of gravity, where similar reductions of the field equations naturally arise.
Moreover, the explicit reconstruction of the metric from the reduced system suggests possible applications in numerical relativity, such as the systematic construction of consistent stationary or quasi-stationary initial data.

Several open problems remain.
A natural extension is to generalize the framework beyond strictly stationary configurations to fully dynamical settings.
In addition, the stability properties of the solutions obtained here-including those with nontrivial topology-have not been analyzed and deserve further study.

Finally, although the ansatz
\begin{equation}
\Sigma(r,x)=P(x)\,\varSigma\bigl(\sigma(r)+Q(x)\bigr)   \nonumber
\end{equation}
is already quite general, it is conceivable that a more flexible functional form could render the right-hand side of Eq.~\eqref{R12=0} integrable under weaker assumptions.
Exploring such generalizations constitutes an interesting direction for future work.

\section*{Acknowledgments}
H.-C.~Kim (RS-2023-00208047) and W.~Lee (RS-2022-NR075087, CQUeST: RS-2020-NR049598)
were supported by Basic Science Research Program through
the National Research Foundation of Korea funded by the Ministry of Education.
We are grateful to Gungwon Kang, Yoonbai Kim, and Yun Soo Myung for helpful
comments, and thank Wontae Kim and Stefano Scopel for their hospitality during our visit to
the Workshop on Cosmology and Quantum Spacetime (CQUeST 2025).

\appendix

\section{Solving angular independence} \label{App1}
Let us solve the angular independence condition of the energy density in Eq.~\eqref{angular eq}
\begin{equation}
\label{AppA: ang-indep}
\partial_z^2 \log \c P = -\beta \c P \,.
\end{equation}

\subsection{$\beta=0$ case}

Before try to solve the most general case, we get a specific solution for the $\beta =0$ case first.
In this case, the differential equation present a solution,
\begin{equation}
\label{cP:special}
\c P (z) = P_0 e^{c_1 z} \,,
\end{equation}
where $c_1$ and $P_0$ are integration constants.
Here, $\frac{d}{dz} = (x^2-1)\frac{d}{dx}$ which gives $z = \frac12 \log \frac{1-x}{1+x}$.
The function $Q$ becomes
\begin{equation}
Q= a^2 \left(\int P dx\right)^2= a^2 \left(\int \c P dz\right)^2
	 = \frac{a^2}{c_1^2}\left(q_0+ \c P\right)^2  \,. \nonumber
\end{equation}

Now, we introduce a new coordinates $\chi$, by using Eq.~\eqref{z cP},  which satisfies
\begin{equation}
\label{App1:dchi dz}
d\chi  =  \sqrt{P}d \theta = \sqrt{\frac{\c P}{1-x^2} } \frac{dx}{-\sqrt{1-x^2}} =  \sqrt{\c P} dz
\quad \rightarrow \quad \chi -\chi_0 = \frac{2}{c_1} \sqrt{\c P} =\frac{2\sqrt{P_0}}{c_1} e^{c_1z/2} \,.
\end{equation}
Then, expressing $g_{\phi \phi} = \c P$ in terms of $\chi$,
\begin{equation}
\label{App1:g33}
g_{\phi\phi} =P\sin^2\theta  = \c P = P_0 e^{c_1 z} =  \frac{c_1^2}{4} (\chi-\chi_0)^2 \,.
\end{equation}
Therefore, we  find that the angular metric becomes
\begin{equation}
\label{cylindrical}
d\Omega_{(2)}^2 = \c P(dz^2+ d\phi^2) = d\chi^2 + \alpha^2 \chi^2 d\phi^2 \,; \qquad \alpha^2 = \frac{c_1^2}{4} \,,
\end{equation}
where we set $\chi_0 =0$ by translating $\chi$ coordinate without loss of generality.
The $Q$ function is
\begin{equation}
\label{Q:flat}
Q(\chi) = \frac{a^2}{4\alpha^2}  \left(q_0+ \alpha^2 (\chi - \chi_0)^2\right)^2 \,.
\end{equation}

\subsection{general case}
Now, we solve the angle-independence~\eqref{angular eq} in a general way.
We use the following trick:
Introducing $w \equiv \c P_{,z}/\c P$, we have
\begin{equation}
\label{App2:1}
w_{,z} = -\beta \c P, \qquad \c P_{,z} = w \c P \,.
\end{equation}
Substituting the first equation to the second, with $\beta \neq 0$,
\begin{equation}
\frac{d}{dz}\left(\frac{w_{,z}}{\beta}\right) = w \frac{w_{,z}}{\beta} \quad
\rightarrow \quad w_{,zz} = w w_{,z}= \frac12 \frac{d w^2}{dz} \,. \nonumber
\end{equation}
For a general case with $w_{,z} \neq 0$, we integrate the equation and get
\begin{equation}
\frac{dw}{dz} = \frac12 w^2 + C_1 \,. \nonumber
\end{equation}
There are three different solutions to this equation depending on the signature of $C_1$.
At present, we begin with the assumption $C_1 \equiv- k^2/2 <0$ and deal the other cases later.
Integrating the equation, we get
\begin{equation}
w(z) = -k \tanh \left( \frac{k}{2}(z-z_0) \right) \quad \rightarrow \quad
\beta \c P \equiv -\frac{dw(z)}{dz} = \frac{k^2}{2} \mbox{sech}^2 \left( \frac{k}{2}(z-z_0) \right) \,. \nonumber
\end{equation}
Notice that
\begin{equation}
\c P = \frac{k^2(1- \tanh^2())}{2\beta}
	= \frac{1}{2\beta} \left(k^2 - w(z)^2\right) \,. \nonumber
\end{equation}
Let us write the results by using the $\chi$ coordinate.
By using the first part of Eq.~\eqref{App1:dchi dz}, and assuming $\beta > 0$,
\begin{equation}
d\chi \equiv \sqrt{\c P} dz =\sqrt{\c P} \frac{dz}{dw} dw  	= \frac{\sqrt{\c P}}{-\beta \c P} dw
	=\frac{\sqrt{2\beta}}{-\beta} \frac{dw}{\sqrt{k^2 - w^2} } \,. \nonumber
\end{equation}
Integrating, we get
\begin{equation}
(\chi-\chi_0) = - \sqrt{\frac{2}{\beta}} \arctan \frac{w}{\sqrt{k^2-w^2}} \,. \nonumber
\end{equation}
Then,
\begin{equation}
\frac{w}{\sqrt{k^2-w^2}} =  \tan \left[\frac{\pi}{2}-\sqrt{\frac{\beta}{2}} (\chi-\chi_0)\right] = \cot \left[\sqrt{\frac{\beta}{2}} (\chi-\chi_0)\right] \,. \nonumber
\end{equation}
Here, we use $\tan (-x + \pi /2) = \cot x$ by choosing the integration constant appropriately, and
\begin{equation}
\quad \rightarrow \quad w^2 = \cot^2() \times (k^2 - w^2) \quad \rightarrow \quad (1+ \cot^2()) w^2 = k^2 \cot^2() \,. \nonumber
\end{equation}
This gives
\begin{equation}
w = \pm k \cos \left[\sqrt{\frac{\beta}{2}} (\chi-\chi_0)\right] \,. \nonumber
\end{equation}
We may choose $+$ signature without loss of generality.
Then,
\begin{equation}
P\sin^2\theta = \c P =  \frac{k^2}{2\beta} \left(1 - \cos^2() \right) =  \frac{k^2}{2\beta} \sin^2 \left[\sqrt{\frac{\beta}{2}} (\chi-\chi_0)\right]  \,. \nonumber
\end{equation}
Finally, we need to calculate the $Q$ function in Eq.~\eqref{P:dQ}.
The $Q$ function becomes, noting $dw/dz = -\beta \c P$,
\begin{equation}
Q = a^2 \left(\int \c P dz\right)^2 =a^2 \left( q_0' -\frac{1}{\beta}w(z)\right)^2
= \frac{a^2k^2}{\beta^2}\left(q_0 - \cos \left[\sqrt{\frac{\beta}{2}} (\chi-\chi_0)\right]\right)^2  \,, \nonumber
\end{equation}
where $q_0$ is an integration constant.
When $\beta < 0$, we can use analytic continuation to make $\cos \to \cosh$, $\sin \to i \sinh$.

\section{Prove the compatibility of the differential equation} \label{App A}

Let us prove the compatibility of the three differential equations in Eq.~\eqref{angular eq2}.
For notational simplicity, we denote the ${}' = \frac{d}{dz}$ in this appendix.

The first equation in Eq.~\eqref{angular eq2} becomes
\begin{equation}
\label{App1:eq1-1}
\c P'' = \frac{\c P'^2}{\c P} - 2\left(\Delta_2 - \frac{6Q}{\ell^2}\right) P^2 \,.
\end{equation}
From last equation in Eq.~\eqref{angular eq2}, we get
\begin{equation}
\label{App1:eq3-1}
\frac{d}{dz}(4a^2P^2Q) =( Q'^2)' \quad \rightarrow \quad 8a^2P P' Q + 4a^2P^2 Q' = 2Q' Q''
\quad \rightarrow \quad P'  = \frac{Q'}{4a^2 P Q} (Q'' - 2a^2P^2) \,.
\end{equation}
Now, let us set
\begin{equation}
\label{App1:Bz}
B(z) \equiv  \frac{\partial_z \c P \partial_z Q}{\c P^2} +2Q \left[\Delta_2 - \frac{3}{\ell^2} Q\right]
  -2a^2\c P +2\Delta_0 \,.
\end{equation}
Then, the second equation in Eq.~\eqref{angular eq2} becomes $B(z) =0$.
To see if this is a conserved quantity for the dynamics given by the remaining two equations~\eqref{App1:eq1-1} and \eqref{App1:eq3-1}, we differentiate $B(z)$:
\begin{equation}
B'(z) = \frac{d}{dz} \left(\frac{\c P'Q'}{\c P^2}\right) 	+ 2\frac{d}{dz}\left[Q(\Delta_2- \frac{3Q}{\ell^2}) \right]- 2a^2 \c P' \,. \nonumber
\end{equation}
Let us calculate term by term. The first term:
\begin{equation}
\frac{d}{dz} \left(\frac{\c P'Q'}{\c P^2}\right)
=\frac{\c P''Q'}{\c P^2} + \frac{\c P'Q''}{\c P^2}
	-\frac{2\c P'^2Q'}{\c P^3} \,. \nonumber
\end{equation}
The second term:
\begin{equation}
2\frac{d}{dz}\left[Q(\Delta_2- \frac{3Q}{\ell^2}) \right]
=2Q'(\Delta_2- \frac{3Q}{\ell^2}) - 2Q\frac{3Q'}{\ell^2}
= 2Q'\left( \Delta_2 - \frac{6Q}{\ell^2} \right) \,. \nonumber
\end{equation}
Therefore,
\begin{equation}
B' =\frac{\c P''Q'}{\c P^2} + \frac{\c P'Q''}{\c P^2} 	-\frac{2\c P'^2Q'}{\c P^3} 	+2Q'\left( \Delta_2 - \frac{6Q}{\ell^2} \right)
	- 2a^2 \c P'  \,. \nonumber
\end{equation}
Now, we substitute $\c P''$ from Eq.~\eqref{App1:eq1-1}:
\begin{equation}
B'= -\frac{\c P'^2Q'}{\c P^3} + \frac{\c P'Q''}{\c P^2}
	- 2a^2 \c P' 	=\c P'\left(  -\frac{\c P'Q'}{\c P^3} + \frac{Q''}{\c P^2}- 2a^2\right) \,. \nonumber
\end{equation}
Now, we put $\c P'$ from Eq.~\eqref{App1:eq3-1} and $\c P^2 $ from the last equation in Eq.~\eqref{angular eq2},
\begin{equation}
B' =\c P'\left( -\frac{Q'^2}{4a^2\c P^4Q} (Q'' - 2a^2P^2)   + \frac{Q''}{\c P^2}- 2a^2\right)
=\c P'\left( -\frac{1}{\c P^2} (Q'' - 2a^2P^2)   + \frac{Q''}{\c P^2}- 2a^2\right) =0 \,.
\end{equation}
Therefore, we get $B(z) = {\rm constant}$. This constant should determine $\Delta_0$.

Now, we solve the coupled differential equation of the last equation in Eq.~\eqref{angular eq2} and $B(z) =0$.
From the last equation in Eq.~\eqref{angular eq2}, we have $\c P = \varepsilon \frac{Q'}{2a \sqrt{Q}}$, where $\varepsilon = \pm 1$.
Then,
\begin{equation}
\c P' = \varepsilon\left[ \frac{Q''}{2a \sqrt{Q}} - \frac{Q'^2}{4a Q^{3/2}} \right] \,. \nonumber
\end{equation}
Now, the first term in Eq.~\eqref{App1:Bz} becomes
\begin{equation}
\frac{\c P' Q'}{\c P^2} =\varepsilon a \frac{Q}{Q' Q^{3/2}} \left[ 2QQ''- Q'^2 \right] \,.
\end{equation}
Then, $B(z)=0$ becomes
\begin{equation}
\label{App1:eq3}
\frac{1}{Q'} \left[ QQ''- Q'^2 \right]   +\frac{\varepsilon}{a} \sqrt{Q}\left[Q \left(\Delta_2 - \frac{3}{\ell^2} Q\right)
		+\Delta_0 \right] =0 \,.
\end{equation}
Now, we treat $Q'$ as a function of $Q$ by setting $p(Q) = Q'= \frac{dQ}{dz}$.
We use
\begin{equation}
Q'' = \frac{dp}{dz} = \frac{dQ}{dz} \frac{dp}{dQ} = p \frac{dp}{dQ} \,. \nonumber
\end{equation}
Then, the equation~\eqref{App1:eq3} becomes
\begin{equation}
 Q\frac{dp}{dQ}- p= Q^2\frac{d}{dQ}\left(\frac{p}{Q}\right)=
  -\frac{\varepsilon}{a} \sqrt{Q}\left[Q \left(\Delta_2 - \frac{3}{\ell^2} Q\right)
		+\Delta_0 \right]  \,. \nonumber
\end{equation}
Integrating with respect to $Q$, we get
\begin{equation}
\label{App1:pQ}
p = \frac{dQ}{dz} = -\frac{\varepsilon Q}{a} \int dQ \left[\Delta_2Q^{-1/2}  - \frac{3}{\ell^2} Q^{1/2} +\Delta_0 Q^{-3/2}\right]
=-\frac{2\varepsilon}{a} \left[\Delta_2Q^{3/2}  - \frac{1}{\ell^2} Q^{5/2} 	-\Delta_0 Q^{1/2}\right]+ C_1 Q  \,,
\end{equation}
where $C_1$ is an integration constant.
Thus,
\begin{equation}
z-z_0 = \int \frac{dQ}{-\frac{2\varepsilon}{a} \left[\Delta_2Q^{3/2}  - \frac{1}{\ell^2} Q^{5/2}
		-\Delta_0 Q^{1/2}\right]+ C_1 Q}  \,. \nonumber
\end{equation}
The integral is cubic polynomial in $\sqrt{Q}$, so the integral is expressible in terms of elementary functions only in special cases.

\end{document}